\documentclass[a4paper,11pt]{article}
\usepackage{float}
\usepackage[pdftex]{graphicx}
\usepackage{subcaption}
\usepackage[T1]{fontenc}
\usepackage{lmodern}
\usepackage{slashed}
\usepackage[utf8]{inputenc}
\usepackage[english]{babel}
\usepackage{microtype}
\usepackage{cite}
\usepackage{amsmath,amssymb,amsfonts,amsthm}
\usepackage{mathtools,mathrsfs,calligra,aurical}
\usepackage[nottoc,notlot,notlof]{tocbibind}
\usepackage{upgreek}
\usepackage{mathtools}
\numberwithin{equation}{section}
\allowdisplaybreaks
\usepackage[all]{xy}
\usepackage{xcolor}
\usepackage{graphicx}
\graphicspath{{images/}}
\usepackage{geometry}
\usepackage[toc,page]{appendix}
\usepackage{hyperref}
\usepackage[normalem]{ulem}
\hypersetup{colorlinks = true, linkcolor = red, linktocpage = true, citecolor = blue}

\newcommand{\ii}{\mathrm{i}}

\usepackage{bm}
\usepackage{ragged2e}
\usepackage{appendix}
\usepackage{slashed}
\usepackage{bbold}
\usepackage{cancel}

\newcommand{\be}{\begin{equation}}
\newcommand{\bea}{\begin{eqnarray}}

\newcommand{\ee}{\end{equation}}
\newcommand{\eea}{\end{eqnarray}}

\DeclareMathAlphabet{\mathpzc}{OT1}{pzc}{m}{it}

\global\long\def\dd{\mathrm{d}}%
\global\long\def\not#1{\slashed{#1}}%

\global\long\def\ii{\imath}%
\global\long\def\jj{\jmath}%

\global\long\def\bR{\mathbb{R}}%
\global\long\def\bC{\mathbb{C}}%

\global\long\def\hM{\mathscr{M}}%
\global\long\def\tr{\mathrm{Tr}}%
\global\long\def\ri{\mathrm{i}}%

\usepackage{setspace}

\begin{document}

\begin{titlepage}
\begin{flushright}
\par\end{flushright}
\vskip 0.5cm
\begin{center}
\begin{spacing}{2.4}
\textbf{\huge The Instability of Low-Temperature Black Holes in Gauged $\mathcal{N}=8$ Supergravity}\\
\end{spacing}

\vskip 5mm

\vskip 1cm

\large {\bf Andr\'{e}s Anabal\'{o}n}$^{~a}$\footnote{anabalo@gmail.com}, \large {\bf Stefano Maurelli}$^{~b,c}$\footnote{stefano.maurelli@polito.it}, \large {\bf Marcelo Oyarzo}$^{~a,b}$\footnote{moyarzoca1@gmail.com} and \large {\bf Mario Trigiante}$^{~b,c}$\footnote{mario.trigiante@polito.it}

\vskip .5cm 

$^{(a)}${\textit{Departamento de F\'isica, Universidad de Concepci\'on, Casilla, 160-C, Concepci\'on, Chile.}}\\ \vskip .1cm
$^{(b)}${\textit{Politecnico di Torino, Dipartimento di Scienza Applicata e Tecnologia, corso Duca degli Abruzzi 24, 10129 Torino, Italy.}}
\\ \vskip .1cm
$^{(c)}${\textit{INFN, Sezione di Torino, Via P. Giuria 1, 10125 Torino, Italy.}}
\end{center}

\vskip .5cm 
\begin{abstract}
We consider the static planar black hole solutions in the STU model of the gauged $\mathcal{N}=8$ supergravity in four dimensions. We give a straightforward derivation of the equation of state of the purely electric and purely magnetic solutions with four charges. Then we give a simple proof that the determinant of the Hessian of the energy is always negative below some critical finite temperature for the purely electric solutions. We compute the spinodal line for the usual planar Reissner–Nordström solution in four dimensions. Inspired by the magnetic superalgebra we show that the supersymmetric solutions are metastable if the energy is restricted to satisfy the topological twist condition ab initio and it is shifted to be zero on the BPS solutions.
\end{abstract}

\vfill{}
\vspace{1.5cm}
\end{titlepage}

\setcounter{footnote}{0}
\tableofcontents

\section{Introduction}
Black holes in $SO(8)$-gauged $\mathcal{N}=8$ supergravity yield concrete examples of the physics of the low energy limit of M-theory. The four-dimensional theory was originally constructed in \cite{deWit:1982bul} and proven to describe the massless sector in the spontaneous compactification of eleven-dimensional supergravity on $S^7$ \cite{deWit:1986oxb}. The consistent truncation of this theory to the four vectors gauging the Cartan subalgebra of $SO(8)$ is the $\mathcal{N}=2$ $U(1)^4$-STU model, whose bosonic sector consists of the metric, four $U(1)$-vectors, three dilatons and three axions. The static-spherically symmetric black holes in this model were obtained with either four electric or four magnetic charges in \cite{Duff:1999gh}. These black holes were embedded in eleven dimensions and generalized to hyperbolic and planar horizons in \cite{Cvetic:1999xp}. The spinning solution, with two dilatons and two axions set to zero, was constructed in \cite{Chong:2004na}. In \cite{Duff:1999gh}, the BPS limit of the spherical black holes was found to yield naked singularities in the electric case and not to exist in the magnetic one. When the four $U(1)$-vectors are equal to one another, this theory reduces to the minimal gauged $\mathcal{N}=2$ supergravity, with the bosonic sector corresponding to the Einstein-Maxwell theory with a negative cosmological constant. In this limit, spherical black holes exhibit naked singularities \cite{Romans:1991nq}, while finite-area black holes must have a locally hyperbolic horizon \cite{Caldarelli:1998hg} and it was unknown whether running scalars would allow for spherical and planar black holes. Eventually, the first regular, finite-area spherically symmetric and planar supersymmetric black hole in $\mathcal{N}=8$ supergravity was constructed in \cite{Cacciatori:2009iz}. The $\mathcal{N}=8$ black holes are known to contain different kinds of instabilities which have been studied restricting the number of charges \cite{Cvetic:1999rb, Chamblin:1999tk, Gubser:2000ec}. 

In this paper, we provide a general and simple proof that the four-charge planar electric black holes of gauged $\mathcal{N}=8$ supergravity are unstable when the temperature is low enough. Indeed, there is a finite temperature at which the black holes are no longer equilibrium states in the thermodynamical sense. To this end, we construct the equation of state of these black holes and compute the determinant of the Hessian of the energy showing that, below a certain temperature, it is always negative. This means that there is a spinodal line for these black holes and we construct it explicitly in the pure Einstein-Maxell case.

This result is puzzling for the magnetic supersymmetric black hole. Indeed, from electromagnetic duality, one would expect that the same equation of state would apply in this case by replacing the electric charge squared with the magnetic charge squared. However, this would mean that the supersymmetric black hole is unstable. Following the results of \cite{Hristov:2011ye, Hristov:2011qr} we propose that the Hessian has to be computed on an energy that goes to zero in the BPS limit and that satisfies the topological twist condition ab initio, a condition necessary for the supersymmetric charge to exist asymptotically. Using this mass we find that the black hole is indeed meta-stable. Metastability of supersymmetric black holes should be expected as they are at the boundary of the allowed region of stability.

In the first section of the paper, we present the Lagrangian and our conventions. Then we provide the general non-extremal solutions of \cite{Duff:1999gh, Cvetic:1999xp} written in a slightly different way. Eventually, we derive the BPS limits of the electric and magnetic black holes. In the second section, we compute the equation of state of electric black holes and show that there is a critical temperature at which the Hessian is always negative. There we compute the spinodal line for the usual Reissner–Nordström black hole. Then we analyze the magnetic case, where there are supersymmetric black holes of finite area. We find that if the ensemble of black holes is required to have boundary conditions that allow for the existence of a finite supercharge asymptotically, and the mass is shifted, the Hessian is indeed positive definite. We also discuss the stability of the non-BPS extremal solutions which admit a first-order description in terms of a fake-superpotential \cite{DallAgata:2010ejj,Gnecchi:2012kb,Klemm:2012vm}. There are appendixes with details which are relevant to the calculations provided in the body of the paper.

\global\long\def\ri{\mathrm{i}}%

\section{STU model}
We are interested in studying the dilatonic sub-sector of the STU model, where the complex scalars parameterizing the special K\"ahler
manifold are purely imaginary $z_{i}=\ri e^{-\phi_{i}}$, $i=1,2,3$, which leads to a substantial simplification of the model. The effective action principle, that we consider for practical proposes, is given by
\begin{align}
\mathcal{S} & =\int\dd^{4}x\sqrt{-g}\left(\frac{R}{2}-\frac{1}{4}\sum_{i}\partial_{\mu}\phi_{i}\partial^{\mu}\phi_{i}-\frac{1}{4}\sum_{\Lambda}Y_{\Lambda}F_{\mu\nu}^{\Lambda}F^{\Lambda\mu\nu}+\frac{1}{L^{2}}\sum_{i}\cosh\phi_{i}\right)\,.\label{action dilatonic STU}
\end{align}
The field equations coming from the action principle (\ref{action dilatonic STU})
are given by 

\begin{align*}
\dd(Y_{\Lambda}\star F^{\Lambda}) & =0\,,\hspace{1cm}\Lambda=1,\dots,4\\
\frac{1}{2}\Box\phi_{i}+\frac{1}{L^{2}}\sinh\phi_{i}-\frac{1}{4}\sum_{\Lambda=1}^{4}\frac{\partial Y_{\Lambda}}{\partial\phi_{i}}F_{\mu\nu}^{\Lambda}F^{\Lambda\mu\nu} & =0\,,\hspace{1cm}i=1,2,3\,,\\
R_{\mu\nu}-\frac{1}{2}g_{\mu\nu}R & =T_{\mu\nu}^{(\phi)}+T_{\mu\nu}^{(F)}\,,
\end{align*}
with
\begin{align*}
T_{\mu\nu}^{(\phi)} & =\frac{1}{2}\sum_{i=1}^{3}\left[\partial_{\mu}\phi_{i}\partial_{\nu}\phi^{i}-g_{\mu\nu}\left(\frac{1}{2}(\partial\phi_{i})^{2}-\frac{2}{L^{2}}\cosh\phi_{i}\right)\right]\,,\\
T_{\mu\nu}^{(F)} & =\sum_{\Lambda=1}^{4}Y_{\Lambda}\left(F_{\mu\rho}^{\Lambda}F^{\Lambda}{}_{\nu}{}^{\rho}-\frac{1}{4}g_{\mu\nu}F_{\rho\sigma}^{\Lambda}F^{\Lambda\rho\sigma}\right)\,.
\end{align*}
They partially reproduce the field equations of the dilatonic sector of the STU
model. It is required to impose also the following equations
\begin{align}
e^{2\phi_{1}}F^{2}\wedge F^{3}-F^{1}\wedge F^{4} & =0\,,\nonumber \\
e^{2\phi_{2}}F^{1}\wedge F^{3}-F^{2}\wedge F^{4} & =0\,,\\
e^{2\phi_{2}}F^{2}\wedge F^{2}-F^{3}\wedge F^{4} & =0\,,\nonumber 
\end{align}
coming from the truncation of the axion fields to zero.
These constraints are automatically fulfilled for purely electric and
purely magnetic configurations. The quantities $Y_{\Lambda}$ depends
on the scalar fields and are minus the diagonal components of the
generalized couplings $\mathcal{I}_{\Lambda\Sigma}$ in the purely dilatonic
sector, see Appendix \ref{Appendix STU model} for further details. They are given explicitly
by
\begin{align*}
Y_{i} & =e^{\phi_{1}+\phi_{2}+\phi_{3}-2\phi_{i}}\,,\qquad Y_{4}=e^{-\phi_{1}-\phi_{2}-\phi_{3}}\,,\qquad i=1,2,3\,.
\end{align*}

Reducing to the dilaton fields, some simplifications occur in the supersymmetry transformations: the K\"ahler connection vanishes $\mathcal{Q}_{\mu}=0$,
the first four components of the $U(1)$-section are real $L^{\Lambda}\in\bR$,
and the superpotential is real $\mathcal{W}\in\bR$. Hence the Killing
spinor equation coming from the supersymmetry variation of the gravitino reduces to
\begin{align}
\delta\Psi_{\mu}^{A}\dd x^{\mu} & \equiv\mathfrak{D}^{A}(\epsilon)\label{KS equation chiral}\\
 & \equiv\dd\epsilon^{A}+\frac{1}{4}\omega_{ab}\gamma^{ab}\epsilon^{A}+\frac{1}{2}A^{M}\theta_{M}\varepsilon^{AB}\delta_{BC}\epsilon^{C}+\frac{1}{4}L^{T}\mathcal{I}F_{ab}\gamma^{ab}\gamma\varepsilon^{AB}\epsilon_{B}+\frac{1}{2}\mathcal{W}\gamma\delta^{AB}\epsilon_{B}=0\,, \notag
\end{align}
where we have defined the 1-form Clifford algebra valued $\gamma=\gamma_{a}e^{a}$. The supersymmetry variation for the gaugini are
\begin{align}
\delta\lambda^{iA} & =-\gamma^{\mu}\partial_{\mu}z^{i}\epsilon^{A}+\frac{1}{2}g^{i\overline{\jj}}\overline{f}_{\overline{\jj}}^{\Lambda}\mathcal{I}_{\Lambda\Sigma}F_{ab}^{\Sigma}\gamma^{ab}\varepsilon^{AB}\epsilon_{B}+g^{i\overline{\jj}}\overline{\mathcal{U}}_{\overline{\jj}}^{M}\theta_{M}\delta^{AB}\epsilon_{B}=0\,,
\end{align}
where in this cases the quantities $z^{i},\overline{f}_{\overline{\jj}}^{\Lambda}$
and $\overline{U}_{\overline{\jj}}{}^{M}\theta_{M}$ are purely imaginary
in the dilatonic sector. The quantity $\overline{U}_{\overline{\jj}}{}^{M}$
is the U(1) covariant derivative of the U(1) section $\overline{V}^{M}$
and $\overline{f}_{\overline{\jj}}{}^{\Lambda}$ are the first four
symplectic components of $\overline{U}_{\overline{\jj}}{}^{M}$. The local supersymmetry parameters are chiral spinors satisfying $\gamma_{5}\epsilon_{A}=\epsilon_{A}$ and $\gamma_{5}\epsilon^{A}=-\epsilon^{A}$, and are related to each other through complex conjugation $(\epsilon^{A})^*=\epsilon_A$. $A^{M}$ are the symplectic gauge fields, and the embedding tensor is given by
\begin{align}
\theta_{M} & =\frac{1}{\sqrt{2}L}(1,1,1,1,0,0,0,0)\,.
\end{align}
The supercovariant derivative (\ref{KS equation chiral}) is the object
entering the Dirac bracket between supercharges that we outline
in the next section. Since we are considering real $\gamma$-matrices,
all the coefficients in the above equation are real. We can write
an equation for the Majorana spinor $\psi^{A}=\epsilon^{A}+\epsilon_{A}$
by combining (\ref{KS equation chiral}) with its complex conjugated.
For simplicity, it is useful to define a complex spinor whose real
and imaginary parts are the $SU(2)$ components of the Majorana spinor
$\zeta=\psi^{1}+\ri\psi^{2}$. The Killing spinor equation for the complex Killing spinor is
\begin{align}
\left(\dd+\frac{1}{4}\omega_{ab}\gamma^{ab}-\frac{\ri}{2}A^{M}\theta_{M}-\frac{\ri}{4}L^{T}\mathcal{I}F_{ab}\gamma^{ab}\gamma+\frac{1}{2}\mathcal{W}\gamma\right)\zeta & =0\,.
\end{align}
The same can be done for the gaugino equations that implies the following equations for the complex spinor
\begin{align}
\left(-\gamma^{\mu}\partial_{\mu}z^{i}-\frac{\ri}{2}g^{i\overline{\jj}}\overline{f}_{\overline{\jj}}^{T}\mathcal{I}F_{ab}\gamma^{ab}+g^{i\overline{\jj}}\overline{U}_{\overline{\jj}}{}^{M}\theta_{M}\right)\zeta & =0\,,\label{gaugino equations zeta}
\end{align}
See Appendix \ref{Appendix STU model} for the definitions and the explicit form of the $\gamma$-matrices that we are considering.

\subsection{Dirac bracket between supercharges }

In gauge theories, the conserved charges associated with large gauge transformations are obtained by the integration of a suitable conserved current on a co-dimension 2 spacelike surface, due to the fact that the Hodge dual of the Noether current is a closed form, up to imposing the field equations \cite{Compere:2019qed}. In
\cite{Hristov:2011ye} was developed a method to evaluate the superalgebra
in backgrounds that have an asymptotic Killing spinor for $\mathcal{N}=2$ gauged
supergravity solutions. This prescription was applied to compute
the BPS bound for configurations that asymptote to AdS or mAdS, resolving
the tension between the BPS bound propose in \cite{Kostelecky:1995ei}
and the explicit BPS configurations found by Romans \cite{Romans:1991nq}.
The same analysis was carried out in \cite{Hristov:2011qr} for $D=4$,
$\mathcal{N}=2$ gauged supergravity coupled to matter fields, where it was shown that the 3-form dual to the Noether supercurrent can be written as the exterior derivative of a 2-form up to imposing the field equations. This leads the supercharge to be expressed as an integral of a 2-form over a spacelike surface in the asymptotic region. The Dirac bracket between the supercharges is computed by acting with the supersymmetry transformation
on the supercharge leading to
\begin{align}
\{\mathcal{Q},\mathcal{Q}\} & =-2\int_{\partial\Sigma}\left(\overline{\epsilon}^{A}\gamma\wedge\mathfrak{D}_{A}(\epsilon)-\overline{\epsilon}_{A}\gamma\wedge\mathfrak{D}^{A}(\epsilon)\right)\,.\label{superalgebra bracket}
\end{align}
where $\mathfrak{D}^{A}(\epsilon)$ is the
supercovariant derivative defined in (\ref{KS equation chiral}), and $\mathfrak{D}_{A}(\epsilon)$ is its complex conjugated. We defined the Clifford algebra valued 1-form $\gamma = \gamma_{a} e^a$. We consider
$\mathcal{Q}$ being the product between the Grassmann-odd supercharge and the spinorial parameters $\epsilon^{A}$, i.e. there are no free indices and it is a Grassmann-even quantity. If the superalgebra (\ref{superalgebra bracket}) is computed in a supersymmetric background, the spinor $\epsilon_A$ is chosen to be the Killing spinor of the background, leading to zero in the right-hand side. However, one can consider a configuration that breaks supersymmetry in the bulk but it has an asymptotic Killing spinor defined in the asymptotic region by imposing certain boundary conditions on the gravitino and the gaugini
\begin{equation}
    \delta \Psi^A_\mu = o(1/r^n) \, ,\qquad \delta \lambda^{A i} = o(1/r^{n_i}) \, ,
\end{equation}
where the constants $n,n_i>0$ must be chosen in such a way that the conserved charges in \eqref{superalgebra bracket}  are finite. Configurations having the same boundary condition as a supersymmetric background can be understood as excitations of it. We will compute the superalgebra in backgrounds belonging to the dilatonic sector of the STU model. In that case (\ref{superalgebra bracket}) can be written in terms of the Majorana Killing spinor $\psi^{A}$ as (see Appendix \ref{Appendix Dirac bracket} for a detailed derivation.)
\begin{align}
\{\mathcal{Q},\mathcal{Q}\} & =2\int\overline{\psi}^{A}\gamma_{5}\gamma\wedge\left(\delta_{AB}\dd+\frac{1}{4}\omega_{ab}\gamma^{ab}\delta_{AB}+\frac{1}{2}A^{M}\theta_{M}\varepsilon_{AB}+\frac{1}{4}L^{T}\mathcal{I}F_{ab}\gamma^{ab}\gamma\varepsilon_{AB}+\frac{1}{2}\gamma\mathcal{W}\delta_{AB}\right)\psi^{B} \, .
\label{backet between call Q}
\end{align}

\subsection{Electric dilatonic black holes and their singular supersymmetric limits}

The four charges electric black hole in the dilatonic
sector in the STU model, with a spherical horizon, was found in \cite{Duff:1999gh},
and its generalizations to planar and hyperbolic horizons were constructed
in \cite{Cvetic:1999xp}. Here, we present the solution for arbitrary
horizon geometry controlled by the parameter $k=-1,0,1$ leading to
hyperbolic, planar and spherical horizons. The BPS limit of the
configurations can be analysed in a simple way by introducing a parameter
$q$ through a change of coordinates. The metric reads
\begin{align}
\dd s^{2} & =-\frac{f(r)}{\sqrt{H(r)}}\dd t^{2}+\frac{\sqrt{H(r)}}{f(r)}\dd r^{2}+r^{2}\sqrt{H(r)}\left(\frac{\dd x^{2}}{1-kx^{2}}+(1-kx^{2})\dd y^{2}\right)\,,\\
f(r) & =k+\frac{r^{2}}{L^{2}}H(r)-\frac{m}{r}-\frac{q}{r^{2}}\,,\qquad H(r)=H_{1}H_{2}H_{3}H_{4}\,,\qquad H_{\Lambda}=1+\frac{q_{\Lambda}}{r}\,.\label{general electric background}
\end{align}
The dilatons and the gauge fields are
\begin{align}
\phi_{1} & =\frac{1}{2}\log\left(\frac{H_{2}H_{3}}{H_{1}H_{4}}\right)\,,\qquad\phi_{2}=\frac{1}{2}\log\left(\frac{H_{1}H_{3}}{H_{2}H_{4}}\right)\,,\qquad\phi_{3}=\frac{1}{2}\log\left(\frac{H_{1}H_{2}}{H_{3}H_{4}}\right)\,,\\
A^{\Lambda} & =\left( \frac{Q_{\Lambda}}{\sqrt{2}rH_{\Lambda}} -\mu_{\Lambda}\right)\dd t\,,\qquad Q_{\Lambda}^{2}=q_{\Lambda}^{2}k+q_{\Lambda}m-q\,.
\end{align}
Where $\mu_\Lambda$ is related to the electric chemical potential which is fixed in terms of the rest of the parameters in such a way that the the gauge fields are regular in the Euclidean configuration. A necessary condition, and sufficient in this case, to preserve some amount of supersymmetry is
that the matrices of the gaugino equations (\ref{gaugino equations zeta})
are not invertible. Imposing that the matrices have zero determinant
implies the following relation between the parameters 
\begin{align}
-kq & =\frac{m^{2}}{4}\,.\label{condition BPS electric}
\end{align}
The BPS configurations exist only in the spherical and planar case, even though they are naked singularities, they have a Killing spinor defined on the geometry. For the hyperbolic case the condition (\ref{condition BPS electric}) leads to purely imaginary gauge fields, which only can make sense by Wick rotating the time coordinate leading to a Euclidean configuration without a Lorentzian limit. 

The Majorana Killing spinor for the planar BPS configuration can be found and is given by
\begin{align}
\psi_{\mathrm{pl}}^{A}(r) & =\frac{f_{\mathrm{pl}}^{1/4}(r)}{2H^{1/8}(r)}(\alpha_{\mathrm{pl}}(r)-\beta_{\mathrm{pl}}(r)\gamma_{1})(\delta_{AB}+\varepsilon_{AB}\gamma_{0})\psi_{0}^{B}\,,
\end{align}
where $\psi_{0}^{B}$ is Majorana constant spinor and the relevant
functions are 
\begin{align}
\alpha_{\mathrm{pl}}(r)=\left(1-\frac{\sqrt{-q}}{rf_{\mathrm{pl}}^{1/2}}\right)^{1/2}\,,\qquad\beta_{\mathrm{pl}}(r)=\left(1+\frac{\sqrt{-q}}{rf_{\mathrm{pl}}^{1/2}}\right)^{1/2}\,,\qquad f_{\mathrm{pl}}(r)=\frac{r^{2}}{L^{2}}H-\frac{q}{r^{3}}\,.
\end{align}
The projector $\delta_{AB}+\varepsilon_{AB}\gamma_{0}$ has matrix
rank equal 2, which implies that planar BPS configuration preserves
four real supercharges. Note also that the Killing spinor diverges
at the curvature singularities of the manifold which are located at
$H(r)=0$. 

The Majorana Killing spinor for the spherical BPS configuration is explicitly given by 
\begin{align}
\psi_{\mathrm{sp}}^{A}(t,r,x,y) & =\frac{f_{\mathrm{sp}}^{1/4}(r)}{2H^{1/8}(r)}e^{\frac{\ri t}{2L}}e^{\frac{\ri}{2}\gamma_{012}\arccos x}e^{-\frac{1}{2}y\gamma_{23}}(\alpha_{\mathrm{sp}}(r)+\beta_{\mathrm{sp}}(r)\gamma_{1})(\delta_{AB}+\varepsilon_{AB}\gamma_{0})\psi_{0}^{B}\,,
\end{align}
where the radial functions are
\begin{align}
\alpha_{\mathrm{sp}}(r) & =\left(1+\frac{1-\frac{m}{2r}}{f_{\mathrm{sp}}^{1/2}}\right)^{1/2}\,,\qquad\beta_{\mathrm{sp}}(r)=\left(1-\frac{1-\frac{m}{2r}}{f_{\mathrm{sp}}^{1/2}}\right)^{1/2}\,,\qquad f_{\mathrm{sp}}(r)=\frac{r^{2}}{L^{2}}H+\left(1-\frac{m}{2r}\right)^{2}\,.
\end{align}
This background preserves four real supercharges, hence it is 1/2 BPS, and the Killing
spinor diverges at the singularity. The spinor depends on all the coordinates, which is also the case in the purely AdS background.
\subsection{Magnetic dilatonic black holes}

The first example of, asymptotically globally
AdS spacetime, supersymmetric static black holes was given by Cacciatori and Klemm \cite{Cacciatori:2009iz}
by considering extremal magnetically charged black holes. They also considered hyperbolic and planar
horizon topology. The non-extremal version of the spherical black
holes with an arbitrary number of vector multiplets and FI terms was
constructed in \cite{Toldo:2012ec}. In \cite{Klemm:2012vm} a family of non-extremal solutions was constructed which contain, in certain limits, the solutions of \cite{Cacciatori:2009iz} and of 
\cite{Cvetic:1999xp,Duff:1999gh}. 
Here we present the dilatonic, magnetic black hole  configurations
in a slightly different parametrization which allows us to straightforwardly connect, through a suitable BPS limit, the magnetic version of the four-dimensional black holes constructed in
\cite{Cvetic:1999xp,Duff:1999gh} to the BPS configurations of \cite{Cacciatori:2009iz}, when the theory allows an embedding in the maximal gauged supergravity. 

The metric of the configuration is
\begin{align}
\dd s^{2} & =-\frac{f(r)}{\sqrt{H(r)}}\dd t^{2}+\frac{\sqrt{H(r)}}{f(r)}\dd r^{2}+r^{2}\sqrt{H(r)}\left(\frac{\dd x^{2}}{1-kx^{2}}+(1-kx^{2})\dd y^{2}\right)\,,\label{general magnetic configuration}\\
f(r) & =\frac{r^{2}}{L^{2}}H(r)+k-\frac{m}{r}-\frac{q}{r^{2}}\,,\qquad H(r)=H_{1}H_{2}H_{3}H_{4}\,,\qquad H_{\Lambda}=1+\frac{q_{\Lambda}}{r}\,,
\end{align}
and the matter fields are given by
\begin{align}
\phi_{1} & =-\frac{1}{2}\log\left(\frac{H_{2}H_{3}}{H_{1}H_{4}}\right)\,,\qquad\phi_{2}=-\frac{1}{2}\log\left(\frac{H_{1}H_{3}}{H_{2}H_{4}}\right)\,,\qquad\phi_{3}=-\frac{1}{2}\log\left(\frac{H_{1}H_{2}}{H_{3}H_{4}}\right)\,,\nonumber \\
A^{\Lambda} & =\frac{1}{\sqrt{2}}P_{\Lambda}x\dd y\,,\qquad P_{\Lambda}^{2}=q_{\Lambda}^{2}k+q_{\Lambda}m-q\,,
\end{align}
Note that the magnetic charges $P_{\Lambda}$ are related to the rest
of the parameters in order to satisfy the field equations. This configuration
corresponds to the magnetic black holes constructed in \cite{Duff:1999gh}
with different topologies for the horizon. The parameter $q$ allow
to analyze the BPS equations in a simple way, and it was introduced
through a diffeomorphism by shifting the radial coordinate.

The vanishing of the determinant of matrices entering in the gaugini variations (\ref{gaugino equations zeta})
implies a relation between the parameters that can be
established in a simple way in terms of the variable $p_{\Lambda}$
related to the $q_{\Lambda}$ as
\begin{align}
\left(\begin{array}{c}
q_{1}\\
q_{2}\\
q_{3}\\
q_{4}
\end{array}\right) & =\left(\begin{array}{cccc}
1 & 1 & 1 & 1\\
1 & -1 & -1 & 1\\
1 & -1 & 1 & -1\\
1 & 1 & -1 & -1
\end{array}\right)\left(\begin{array}{c}
p_{1}\\
p_{2}\\
p_{3}\\
p_{4}
\end{array}\right)\,.
\end{align}
The supersymmetry conditions are given by
\begin{align}
m & =-2kp_{1}+\frac{8}{L^{2}}p_{2}p_{3}p_{4}\,,\label{BPS cond m magnetic}\\
q & =-\frac{1}{4}k^{2}L^{2}+k(-p_{1}^{2}+p_{2}^{2}+p_{3}^{2}+p_{4}^{2})-\frac{4}{L^{2}}(p_{3}^{2}p_{4}^{2}+p_{2}^{2}p_{3}^{2}+p_{2}^{2}p_{4}^{2}-2p_{1}p_{2}p_{3}p_{4})\,.\label{BPS cond q magnetic}
\end{align}
These are sufficient conditions to have a non-trivial Killing spinor
satisfying (\ref{KS equation chiral}). In this limit, the magnetic charges and the metric function can be expressed as follows:
\begin{align}
\left(\begin{array}{c}
P_{1}\\
P_{2}\\
P_{3}\\
P_{4}
\end{array}\right) & =\frac{1}{2L}\left(\begin{array}{cccc}
1 & 1 & 1 & 1\\
1 & -1 & -1 & 1\\
1 & -1 & 1 & -1\\
1 & 1 & -1 & -1
\end{array}\right)\left(\begin{array}{c}
kL^{2}\\
4p_{3}p_{4}\\
4p_{2}p_{3}\\
4p_{2}p_{4}
\end{array}\right)\,,\label{POmp}\\
f(r) & =\frac{1}{L^{2}r^{2}}\left(\frac{k}{2}L^{2}+(r+p_{1})^{2}-p_{2}^{2}-p_{3}^{2}-p_{4}^{2}\right)^{2}\,,
\end{align}
where $p_\Lambda$ are real provided:\footnote{If this condition is not met, we need to change the last three signs in the last column of the matrix on the right-hand side of eq. \eqref{POmp}.}
$$(P_1+P_2- k L)(P_1+P_3- k L)(P_2+P_3-k L)<0\,.$$
The metric function factorizes in a perfect square, then if there
is any horizon, it will be an extremal horizon. For the BPS configuration
it follows that
\begin{align}
\sum_{\Lambda}P_{\Lambda}=2kL\quad \Longleftrightarrow \quad \frac{1}{\sqrt{2}}\Gamma^{M}\theta_{M} & =k\,,\label{topological twist condition}
\end{align}
which was recast in a symplectic-invariant way by using $\Gamma^{M}$ that is the symplectic vector of the topological charges $\Gamma^{M}=(P^{\Lambda},Q_{\Lambda})$, and $Q_{\Lambda}$ are the electric charges that in the present case
are zero. The condition (\ref{topological twist condition}) is known
as the topological twist condition, and it was shown in \cite{DallAgata:2010ejj}
that (\ref{topological twist condition}) is a necessarily condition
to have a BPS static configuration with dyonic topological charges
in $D=4$, $\mathcal{N}=2$ supergravity with vector multiples and FI
terms. 

The configurations satisfying the BPS conditions (\ref{BPS cond m magnetic}) and (\ref{BPS cond q magnetic}) have a well-defined Majorana Killing spinor given by
\begin{align}
\psi^{A}(r) & =\frac{f(r)^{1/4}}{H(r)^{1/8}}\frac{1}{4}(1+\gamma_{1})(\delta_{AB}-\varepsilon_{AB}\gamma_{23})\psi_{0}^{B}\,,\label{Killing spinor magnetic black holes}
\end{align}
where $\psi_{0}^{B}$ is a constant Majorana spinor. The projector $(1+\gamma_{1})(\delta_{AB}-\varepsilon_{AB}\gamma_{23})$ rule out 6 of the 8 components of the doublet $\psi_{0}^{B}$. Hence, the magnetic BPS black holes for any geometry of the horizon have 2 real supercharges, which corresponds to 1/4 of the total supersymmetry.

\section{Thermodynamic analysis for planar configurations}

In this section, we will analyze the thermodynamic stability of the planar black holes in both the electric and magnetic cases. In this case, the equation of state can be written analytically in terms of the horizon ``area-density'', given by
\begin{align}
A & =r_{+}^{2}\sqrt{H(r_{+})}\,. \label{horizon area definition}
\end{align}
From now on, with a slight abuse of notation, we will refer to it as the horizon area. In our units the length dimension of $A$ is 2 and the length dimension of the electric charges
$Q_{\Lambda}$ is 1.
\subsection{Stability of electric black holes}
Considering the electric configurations with arbitrary parameters
given in (\ref{general electric background}), we can solve the parameter
$m$, which up to a numerical factor corresponds to the boundary term given by the on-shell value of the Hamiltonian density  \cite{Anabalon:2014fla, Anabalon:2012sn},
in terms of the horizon area and the electric charges $Q_{\Lambda}$,

\begin{align}
E\equiv m & =\frac{1}{A^{1/2}}\prod_{\Lambda=1}^{4}\left(\frac{A^{2}}{L^{2}}+Q_{\Lambda}^{2}\right)^{1/4}\,.\label{energy density electric black holes}
\end{align}
The energy density given in (\ref{energy density electric black holes})
reproduces the Hawking temperature, up to a numerical factor, by computing
its derivative with respect to the horizon area 
\begin{align}
T_{H} & =\frac{m}{4\pi A}\left[\frac{A^{2}}{L^{2}}\sum_{\Lambda}\left(\frac{A^{2}}{L^{2}}+Q_{\Lambda}^{2}\right)^{-1}-1\right]=\frac{1}{2\pi}\frac{\partial E}{\partial A}\,.
\end{align}
For the electric black holes, the regularity of the gauge fields at
the horizon fixes the chemical potentials $\mu_{\Lambda}$ in (\ref{general electric background}).
This can be also reproduced by the computation of the partial derivative
of the energy (\ref{energy density electric black holes}) with respect
to the electric charges
\begin{align}
\mu_{\Lambda} & =\frac{Q_{\Lambda}}{\sqrt{2}r_{+}H_{\Lambda}(r_{+})}=\frac{mQ_{\Lambda}}{\sqrt{2}(\frac{A^{2}}{L^{2}}+Q_{\Lambda}^{2})}=\frac{1}{\sqrt{2}}\frac{\partial E}{\partial Q_{\Lambda}}\,.
\end{align}
Then, we can construct the first law for the electric black holes
\begin{align}
\frac{1}{8\pi}\delta E & =T_{H}\delta S+\frac{1}{8\pi\sqrt{2}}\sum_{\Lambda}\mu_{\Lambda}\delta Q_{\Lambda}\,,
\end{align}
where the factors can be reabsorbed in the definition of the energy
density $E$ and the electric charges $Q_{\Lambda}$. 

Having an expression for the energy density as a function of the physical
charges and the entropy of the electric black holes, we can analyse the stability of the system by studying the Hessian matrix
\begin{align}
\mathcal{H}_{ab} & =\frac{\partial E}{\partial l^{a}\partial l^{b}}\,,\qquad l^{a}=(A/L,Q_{\Lambda})\,.\label{hessian electric}
\end{align}
which is a $5\times5$ symmetric matrix, and hence there is a limitation to find the eigenvalues in a closed form for a generic value of the parameters. However, the determinant of the Hessian can be written in a simple form in terms of the temperature as follows
\begin{align}
\det\mathcal{H} & =\frac{\pi L^{2}}{4A^{3}}T_{H}-\frac{1}{16E^{3}L^{4}}\sum_{\Lambda}Q_{\Lambda}^{2}\,.
\end{align}
Clearly, for all the extremal black holes, i.e. $T_{H}=0$, the Hessian
has at least one negative eigenvalue, indicating an instability. Consequently, all the extremal electrically charged black holes in this ensemble are thermodynamically unstable. Furthermore, even above extremality, there is a finite gap for which these black holes are all unstable.  

Now we will specialize in the computation for electrically charged Reissner–Nordström
black hole with a planar horizon, which is obtained by setting $q_{\Lambda}=0$
and defining $Q^{2}=-q$. The metric function and gauge fields reduce
to its standard form
\begin{align}
f(r) & =\frac{r^{2}}{L^{2}}-\frac{m}{r}+\frac{Q^{2}}{r^{2}}\,,\qquad A^{\Lambda}=\left(\frac{Q}{\sqrt{2}r}-\mu_{\Lambda}\right)\dd t\,,
\end{align}
and the temperature of the black hole becomes
\begin{align}
T_{H} & =\frac{3A^{2}-L^{2}Q^{2}}{4\pi A^{3/2}L^{2}}\,.
\end{align}
In this case, it is possible to compute the eigenvalues of the Hessian
matrix (\ref{hessian electric}) which are given by
\begin{align}
\lambda & =\frac{A^{2}-L^{2}Q^{2}}{2\sqrt{A}(A^{2}+L^{2}Q^{2})}\,,\\
\lambda_{\pm} & =\frac{1}{8A^{1/2}}\left(5+\frac{3L^{2}Q^{2}}{A^{2}}\pm\sqrt{1+22L^{2}\frac{Q^{2}}{A^{2}}+9L^{4}\frac{Q^{4}}{A^{4}}}\right)\,,
\end{align}
where $\lambda$ has multiplicity three. The eigenvalues $\lambda_{\pm}$ are always positive for any value of $A>0$ and $Q\in\bR$, while the triple eigenvalue $\lambda$ is positive for large black holes compared with the AdS radius and the electric charge, namely $A>L |Q|$. Whence the spinodal line, which separates the stable from the unstable region, is located at $A=L|Q|$. Consistently with our previous discussion, the extremal black holes are obtained at $A=L|Q|/\sqrt{3}<L |Q|$, namely outside the stability region. 
\\
\subsection{Stability of magnetic planar black holes}

The planar magnetic black holes presented in (\ref{general magnetic configuration})
have a well-defined BPS limit that generically represents extremal
BPS black holes with a globally defined Killing spinor (\ref{Killing spinor magnetic black holes}).
One can notice that for the magnetic black holes (\ref{general magnetic configuration}),
it is also possible to solve the integration constant $m$ in terms
of the horizon area $A$, defined in (\ref{horizon area definition}),
and the magnetic charges $P_{\Lambda}$ as 
\begin{align}
m & =\frac{1}{A^{1/2}}\prod_{\Lambda}\left(\frac{A^{2}}{L^{2}}+P_{\Lambda}^{2}\right)^{1/4}\,.\label{m as function A Ps}
\end{align}
Then, if $m$ is identified with the energy density of the configurations,
we can run the same argument that we outline for the electric black
holes and conclude that the extremal BPS black holes are unstable.
This is in tension with the fact that the magnetic planar BPS black
holes are vacuum states of the theory, and therefore are believed
to be stable. In what follows, we will show that the configurations
(\ref{general magnetic configuration}) with $k=0$ asymptote to the
BPS configurations, in the sense that they admit an asymptotic Killing
spinor which leads to finite conserved charges, if and only if the
topological twist condition (\ref{topological twist condition}) is
satisfied. Then, we compute the Dirac bracket between the supercharges
for the asymptotic Killing spinor, showing that the quantity that
we would like to identify with the energy density should vanish in
the BPS limit. We will take this fact into account to propose an energy
density that leads to a semi-positive defined Hessian matrix on backgrounds
that satisfy the topological twist condition imposed at the beginning
of the analysis.

To simplify the analysis we consider the supersymmetry transformation
for the complex spinors, and denote the complex gravitino and complex
gaugini as the chiral one but erasing the SU(2) index. The leading
order of the gaugini equations expanded at $r\to\infty$ goes as $1/r$
and is a matrix equation that can be solved by imposing the following
projector 
\begin{align}
\zeta_{\infty} & =\frac{1}{2}(1+\gamma_{1})\chi_{\infty}\,,\label{projector of the asymptotic spinor}
\end{align}
then the subleading term of the complex gaugini equations read 
\begin{align}
\delta\lambda^{i} & =\frac{1}{2r^{2}}\left[\frac{(P_{i}^{2}-P_{4}^{2})^{2}-(P_{j}^{2}-P_{k}^{2})^{2}}{2Lm^{2}}+\sum_{\Lambda}\Omega_{i\Lambda}P_{\Lambda}\ri\gamma_{23}\right]\zeta_{\infty}+o(r^{-3})\:,\qquad i\neq j\neq k\neq4\,,\\
\Omega_{i\Lambda} & =\left(\begin{array}{cccc}
1 & -1 & -1 & 1\\
1 & -1 & 1 & -1\\
1 & 1 & -1 & -1
\end{array}\right)\,.\label{def Omega i Lambda}
\end{align}
It is interesting to notice that the matrix in bracket in the subleading
term is invertible unless the $P_{\Lambda}$ and $m$ satisfy the
BPS conditions; the same will happen for the subleading terms in the
gravitino equations. Using the projector (\ref{projector of the asymptotic spinor})
the gravitino equations for the complex spinor in the asymptotic region
are given by
\begin{align}
\delta\Psi_{t} & =\partial_{t}\zeta_{\infty}+\frac{\ri}{8Lr}\sum_{\Lambda}P_{\Lambda}\gamma_{023}\zeta_{\infty}+o(r^{-2})\,,\\
\delta\Psi_{r} & =\partial_{r}\zeta_{\infty}-\frac{1}{2r}\zeta_{\infty}+o(r^{-2})\,,\\
\delta\Psi_{x} & =\partial_{x}\zeta_{\infty}+\frac{\ri}{8r}\sum_{\Lambda}P_{\Lambda}\gamma_{3}\zeta_{\infty}+o(r^{-2})\,,\\
\delta\Psi_{y} & =\partial_{y}\zeta_{\infty}-\frac{\ri}{4}x\sum_{\Lambda}P_{\Lambda}\zeta_{\infty}+\frac{\ri}{8r}\sum_{\Lambda}P_{\Lambda}\gamma_{2}\zeta_{\infty}+o(r^{-2})\,.\label{eq asymptoty psi y}
\end{align}
Note that the second term in the equation (\ref{eq asymptoty psi y})
is leading in the expansion on $r$, thus, a necessarily condition
on the background to have asymptotic Killing spinors is that the topological
twist condition (\ref{topological twist condition}) must be fulfilled.
Solving the leading order of the above equations equal zero, and going
back to the Majorana spinor, we find that the asymptotic Killing spinor
on a background that fulfills the topological twist condition is given
by 
\begin{align}
\psi_{\infty}^{A}(r) & =\frac{1}{2}r^{1/2}(1+\gamma_{1})\psi_{0}^{A}+o(r^{-1/2})\equiv r^{1/2}\mathbb{P}_{AB}\psi_{0}^{B}\,,\label{asymptotic KS solution}
\end{align}
where $\psi_{0}^{A}$ is a constant doublet of Majorana spinors. Observe
that the radial dependency agrees with both the expansion at infinity
and the projection $1+\gamma_{1}$ of the Killing spinor of the BPS
background given in (\ref{Killing spinor magnetic black holes}),
and there are no further projections of the asymptotic Killing spinor.
Therefore, the asymptotic spinor (\ref{asymptotic KS solution}) has
4 independent real components which represent an enhancement of the
2 independent real components with respect to the global Killing spinor
defined in the vacuum (\ref{Killing spinor magnetic black holes}).
Indeed we can assert that the asymptotic Killing spinor can be split
into two independent spinors obtained by projecting (\ref{asymptotic KS solution})
with the projector
\begin{align}
\mathbb{P}_{AB}^{\pm} & \equiv\frac{1}{4}(1+\gamma_{1})(\delta_{AB}\pm\varepsilon_{AB}\gamma_{23})\,.\label{projector plus minus}
\end{align}

Computing the right-hand-side of the algebra (\ref{superalgebra bracket})
on the background (\ref{general magnetic configuration}) satisfying
the topological twist condition for the asymptotic Killing spinor
(\ref{asymptotic KS solution}) we get the following result
\begin{align}
\{\mathcal{Q},\mathcal{Q}\} & =-\frac{1}{2}\int_{\partial\Sigma}\sum_{\Lambda}H^{1/4}\left[\left(-\frac{\ri r^{2}f^{1/2}}{H_{\Lambda}}+\frac{\ri r^{3}}{L}H_{\Lambda}\right)\overline{\psi}_{0}^{A}\gamma_{0}\mathbb{P}_{AB}\psi_{0}^{B}-\frac{rP_{\Lambda}}{H_{\Lambda}}\overline{\psi}_{0}^{A}\gamma_{5}\varepsilon_{AB}\mathbb{P}_{BC}\psi_{0}^{C}\right]\dd x\wedge\dd y\,.
\end{align}
Note that the last term would be combined with the first term if the
asymptotic spinor would satisfy the extra projection condition that
we are lacking to reproduce from the asymptotic analysis. We proceed
as follows, let us consider the two independent Killing spinors obtained
by projecting the asymptotic Killing spinor (\ref{asymptotic KS solution})
with (\ref{projector plus minus}) and then computing Dirac bracket
between supercharges twice, one for each independent spinor. To emphasize
this fact we include a subscript in the supercharge $\mathcal{Q}_{\pm}$.
The resulting algebra is the following
\begin{align}
\{\mathcal{Q}_{\pm},\mathcal{Q}_{\pm}\} & =\frac{\ri}{2}\int_{\partial\Sigma}\sum_{\Lambda}H^{1/4}\left(-\frac{r^{2}f^{1/2}}{H_{\Lambda}}+\frac{r^{3}}{L}H_{\Lambda}\pm\frac{rP_{\Lambda}}{H_{\Lambda}}\right)\overline{\psi}_{0}^{A}\gamma^{0}\mathbb{P}_{AB}^{\pm}\psi_{0}^{B}\dd x\wedge\dd y\,,\\
 & =\ri\mathbf{M}_{\pm}\overline{\psi}_{0}^{A}\gamma^{0}\mathbb{P}_{AB}^{\pm}\psi_{0}^{B}\int\dd x\wedge\dd y\,,
\end{align}
where 
\begin{align}
\mathbf{M}_{\pm} & =\frac{L}{2m^{3}}(m^{2}\pm m_{\mathrm{BPS}}^{2})(2m^{2}\pm m_{\mathrm{BPS}}^{2})\,,\\
m_{\mathrm{BPS}}^{2} & \equiv\frac{1}{2\sqrt{2}L}\prod_{i=1}^{3}\left(\sum_{\Lambda}\Omega_{i\Lambda}P_{\Lambda}^{2}\right)^{1/2}\,,\label{m BPS}
\end{align}
and $\Omega_{i\Lambda}$ is defined in (\ref{def Omega i Lambda}).
We will show that indeed, the charges are such that $m_{\mathrm{BPS}}^{2}>0$ in the region of the phase space where the black holes exist. As expected the BPS bound computed
with the spinors that asymptote to the Killing spinor of the background
(\ref{Killing spinor magnetic black holes}), i.e. with the lower
sign, leads to a non-trivial constraint on the parameter $m$
\begin{align}
\mathbf{M}_{-}>0\implies m>m_{\mathrm{BPS}}\,.
\end{align}
While the BPS bound coming from the asymptotic Killing spinor with the upper sign is trivially fulfilled. Now, we go back to the issue of
the definition of the energy density in this configuration.

Note that the derivative of (\ref{m as function A Ps}) with respect
to the horizon area correctly reproduces the Hawking temperature of the black hole. To avoid spoiling the above relation, any reasonable
attempt to define a new thermodynamic energy density can only differ
from (\ref{m as function A Ps}) by the addition of a function of
the magnetic charges. 

For configurations satisfying the topological twist condition, we
propose the following definition of energy density
\begin{align}\label{full energy magnetic}
E & =m(A,P_{\Lambda})-m_{\mathrm{BPS}}(P_{\Lambda})\,,
\end{align}
where $m(A,P_{L})$ is given in (\ref{m as function A Ps}) and $m_{\mathrm{BPS}}$
is given by (\ref{m BPS}). Under these considerations, it is straightforward to prove that the $4\times4$ Hessian matrix is semi-positive defined on extremal configurations.

Now we move to the discussion on the existence of black hole configuration
in the extremal limit that satisfies the topological twist condition.
First of all, observe that the location of the horizon in terms of the
horizon area and the location of singularities in terms of the integration
constant $m$ and $q$ are
\begin{align}
r_{0} & =\frac{A^{2}}{L^{2}m}-\frac{q}{m}\,,\hspace{0.8cm}r_{\Lambda}^{(\text{sing})}=-q_{\Lambda}=-\frac{P_{\Lambda}^{2}}{m}-\frac{q}{m}\,,
\end{align}
respectively. The black hole configuration exist if $r_{0}>r_{\Lambda}^{(\text{sing})}$
which implies that
\begin{align}
\frac{1}{m}\left(\frac{A^{2}}{L^{2}}+P_{\Lambda}^{2}\right) & >0\,.\label{good condition existence}
\end{align}
The existence of a zero of the metric function $f(r_{0})=0$ implies
the existence of a horizon with horizon area $A>0$. The equation
(\ref{good condition existence}) implies that if such a condition is
fulfilled, a horizon automatically covers the singularities.

Extremal configurations have Hawking temperature equal to zero. The Hawking
temperature for the magnetic black hole configurations is obtained
by computing the derivative of the energy (\ref{full energy magnetic})
with respect to the entropy over $8\pi$, leading to
\begin{align}
T_{H} & =\frac{m(A,P_{\Lambda})}{4\pi A}\left[\frac{A^{2}}{L^{2}}\sum_{\Lambda}\left(\frac{A^{2}}{L^{2}}+P_{\Lambda}^{2}\right)^{-1}-1\right]\,.\label{TH magnetic}
\end{align}
Replacing the topological twist condition on (\ref{TH magnetic}),
we find that the right-hand-side gets factorized and consequently
\begin{align}
T_{H}=0\hspace{1cm}\implies\hspace{1cm}\mathrm{Pol}_{+}(A,P_{\Lambda})\mathrm{Pol}_{-}(A,P_{\Lambda})=0\,,
\label{polpm}
\end{align}
where
\begin{align}
\mathrm{Pol}_{\pm}(A,P_{\Lambda}) & =(2\pm1)\frac{A^{4}}{L^{4}}+\frac{A^{2}}{2L^{2}}\sum_{\Lambda}P_{\Lambda}^{2}\mp\prod_{\Lambda}P_{\Lambda}\hspace{0.8cm}\text{satifying}\hspace{0.8cm}\sum_{\Lambda}P_{\Lambda}=0\,.
\end{align}
The greater root for $A$ coming from $\mathrm{Pol}_{-}(A, P_{\Lambda})=0$, which we call $A_{-}$, correctly reproduces the horizon area of BPS black holes. While the greater root for $A$ from $\mathrm{Pol}_{+}(A, P_{\Lambda})=0$, which we call $A_+$, corresponds to the horizon area of non-BPS extremal black holes. In the surface with $T_H=0$ for configurations satisfying the topological twist conditions, there are three magnetic charges as remaining free parameters. The horizon areas $A_{+}$ and $A_{-}$ take values on this space. It is interesting to note that in the region where $A_{-}>0$ we have $A_{+}<0$ and vice versa, therefore, the location where $A_{-}=0$ coincides with $A_{+}=0$. This is depicted in Figure \ref{fig plane X1X2}.
\\
One can interpret this result as follows. For extremal black holes, there are certain boundary conditions that allow the existence of supersymmetric magnetic black holes, and its complementary region will lead to extremal non-BPS black holes. These two essentially different boundary conditions lead to the horizon area $A_-$ and $A_+$, respectively, that can be written in terms of quartic invariant quantities constructed out of the embedding tensor $\theta_M$, the topological charges symplectic vector $\Gamma^M$ and $K_{MNPQ}$ rank-4 completely symmetric tensor of $Sp(8,\bR)$. The explicit form of the horizon area for the extremal BPS black holes is\footnote{For the expression of $A_-$ in terms of ${\rm SL}(2,\mathbb{R})^3$-invariants see \cite{Astesiano:2021hro}. Here we also give the analogous expression for the horizon-area $A_+$, corresponding to new extremal non-BPS solutions.}
    \begin{equation}
        A_{-}^{2} = \frac{3}{2}\frac{I_{2}}{I_{0}}+ \displaystyle{\sqrt{\Bigg(\frac{3}{2}\frac{I_{2}}{I_{0}}\Bigg)^{2} - \frac{1}{4}\frac{I_{4}}{I_{0}}}} \, ,
    \end{equation}
since $I_2<0$ then the area is real an if $I_4<0$. The horizon area for extremal non-BPS black holes are
\begin{equation}
        A_{+}^{2} = \frac{1}{2}\frac{I_{2}}{I_{0}}+ \displaystyle{\sqrt{\Bigg(\frac{1}{2}\frac{I_{2}}{I_{0}}\Bigg)^{2} + \frac{1}{12}\frac{I_{4}}{I_{0}}}} \, .
    \end{equation}
again $I_2<0$ then the above area is real if $I_4>0$. We have defined
\begin{align}
I_{0} & =-2\mathcal{I}_{1}(\theta,\theta,\theta,\theta)=\frac{1}{L^{4}}\,,\\
I_{4} & =-2\mathcal{I}_{1}(\Gamma,\Gamma,\Gamma,\Gamma)=4\prod_{\Lambda}P_{\Lambda}\,,\\
I_{2} & =-\frac{2}{3}\sum_{i=1}^{3}\mathcal{I}_{i}(\Gamma,\theta,\Gamma,\theta)=-\frac{1}{6L^{2}}\sum_{\Lambda}P_{\Lambda}^{2}\,,
\end{align}
where the last relation was obtained provided that $\sum_{\Lambda}P_{\Lambda}=0$, see Appendix \ref{appendix quartic invariant} for the definition of the tensor $\mathcal{I}_i$.

\subsection{First-order description and stable extremal non-BPS solutions}
The BPS solutions discussed above admit a first-order description in terms of gradient-flow equations defined by a suitable black-hole superpotential.
The general form of the latter was found, in the spherical horizon case, in \cite{DallAgata:2010ejj}.
A general discussion of the first-order description of extremal solutions in the STU model was performed in \cite{Klemm:2012vm}.
Using the following standard notation for the spacetime metric:
\begin{equation}
    ds^2=-e^{-2U}\,dt^2+e^{-2U}\,dr^2+e^{2(\psi-U)}\,d^2\Omega\,,
\end{equation}
where $d^2\Omega$ is the metric on the horizon and $U=U(r),\,\psi=\psi(r)$, the superpotential for the BPS case can be written in the form:
\begin{equation}
    \mathscr{W}(U,\psi,z^i,\bar{z}^{\bar{\imath}})=e^{U}\,(\mathcal{Z}{+}i\,e^{2(\psi-U)}\mathcal{W})\,.
\end{equation}
where $\mathcal{Z}\equiv V^T\mathbb{C}\,\Gamma$ is the $\mathcal{N}=2$ central charge and $\mathcal{W}\equiv V^T\,\theta$ is the gauge superpotential.
In our solutions:
\begin{equation}
    U(r)=\frac{1}{2}\log\left(\frac{f(r)}{H(r)^{1/2}}\right)\,,\qquad\psi(r)=\log\left(rf(r)^{1/2}\right)
\end{equation}
The scalar fields satisfy the gradient flow equations: 
\begin{align*}
\frac{\dd z^{i}}{\dd r}=-2\,e^{-2\psi}\,g^{i\overline{\jj}}\partial_{\overline{\jj}}|\mathscr{W}|\,.
\end{align*}
There is a non-BPS branch of extremal solutions whose first-order description was studied in \cite{Gnecchi:2012kb,Klemm:2012vm}.
The fake-superpotential for the dilatonic solutions has the form
\begin{equation}
    \mathscr{W}_{{\tiny \mbox{non-BPS}} }(U,\psi,z^i,\bar{z}^{\bar{\imath}})=e^{U}\,\left(\frac{1}{2}\,\left(\mathcal{Z}+\sum_{\hat{\imath}=1}^3\mathcal{D}_{\hat{\imath}}\mathcal{Z}\right)+i\,e^{2(\psi-U)}\mathcal{W}\right)\,,
\end{equation}
where:
\begin{equation}
    \mathcal{D}_{\hat{\imath}}\mathcal{Z}\equiv e_{\hat{\imath}}^j\,\left(\partial_i+\frac{1}{2}\,\partial_i\mathcal{K}\right)\mathcal{Z}\,,
\end{equation}
are the three matter charges, $e_{\hat{\imath}}^j$ being the inverse vielbein matrix, see Appendix \ref{Appendix STU model} for the relevant definitions related to the special geometry of the model.\par
The expressions for the relevant quantities for these non-BPS solutions are obtained from the corresponding ones derived above for the BPS black holes, upon changing $P_4\rightarrow -P_4$.
The topological twist condition \eqref{topological twist condition}, for instance, for the flat-horizon case, becomes:
\begin{equation}P_1+P_2+P_3-P_4=0\,.\label{tt2}\end{equation}
Just as it happened for the BPS case, the above condition implies a factorization of the expression of the temperature $T_H$. The horizon area now corresponds to a root $A'_-$ given by that of $A_-$ by changing $P_4\rightarrow -P_4$:
    \begin{equation}
        A_{-}^{\prime 2} = \frac{3}{2}\frac{I_{2}}{I_{0}}+ \displaystyle{\sqrt{\Bigg(\frac{3}{2}\frac{I_{2}}{I_{0}}\Bigg)^{2} + \frac{1}{4}\frac{I_{4}}{I_{0}}}} \, ,
    \end{equation}
since $I_2<0$ then the area is real an if $I_4>0$. 
The expression of the Hessian, as a function of the charges, is the same as that of the BPS case. By the same token, we then conclude that 
also these non-BPS extremal solutions, described by first-order gradient flow equations, are stable. 
Stability seems then to be implied by the existence of a first-order description of the solution.\par
By the same token, we also find, for suitable values of the magnetic charges, extremal non-BPS solutions with area $A_+'$ whose expression is obtained from that of $A_+$ by changing $P_4\rightarrow -P_4$:
\begin{equation}
        A_{+}^{\prime 2} = \frac{1}{2}\frac{I_{2}}{I_{0}}+ \displaystyle{\sqrt{\Bigg(\frac{1}{2}\frac{I_{2}}{I_{0}}\Bigg)^{2} - \frac{1}{12}\frac{I_{4}}{I_{0}}}} \, . 
\end{equation}
Reality  of $A'_+$ then requires $I_{4}<0$. We conclude that the condition $I_4<0$ does not uniquely define the BPS configuration. This is to be expected since, in the two cases, the linear conditions on the charges, \eqref{topological twist condition} and \eqref{tt2}, are different.
\begin{center}
\begin{figure}
\begin{centering}
\includegraphics[scale=0.6]{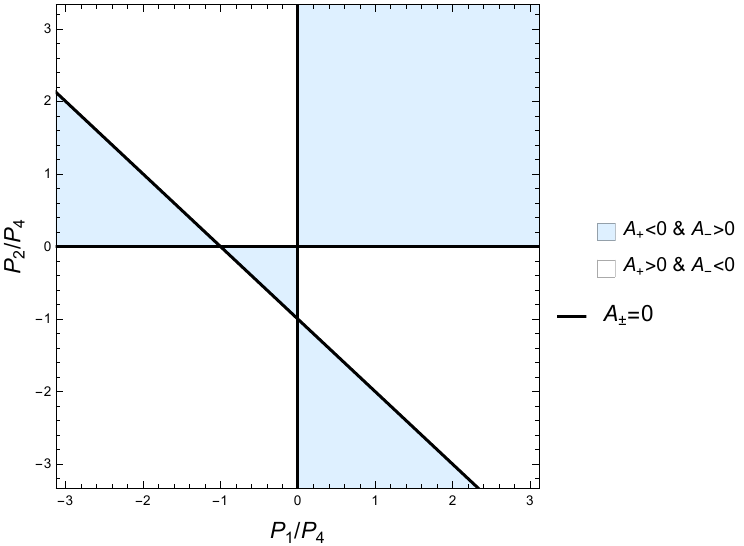}
\par\end{centering}
\caption{Considering $P_3=-P_1-P_2-P_4$, 
 we consider the plane $(P_{1}/P_{4},P_{2}/P_{4})$ where the coloured regions correspond to the existence of susy black holes with $A_{-}>0$. We defined $A_\pm$ as the grater root for $A$ of $\mathrm{Pol}_{\pm}(A,P_{\Lambda})=0$
\label{fig plane X1X2}}
\end{figure}
\par\end{center}
\section{Conclusion}
In this work, we presented a simple argument to prove the thermodynamic instability of extremal planar 4-charges electric black holes in the STU model of the maximal theory in $D=4$. This result constitutes a generalization of the instabilities of electric black holes studied in \cite{Gubser:2000ec}. The argument can be made sharper to conclude that there is a finite critical lower temperature for which electrically charged black holes are unstable, indicating that the mechanics that trigger the instability do not rely on particular features of extremal black holes. We note that indeed it is well known that extremal black holes are unstable under charged perturbations \cite{Gubser:2008px}. However, the phenomenon presented here is different because it does not require extremality and occurs at fixed charges. It seems, in other words, to somehow capture the non-linearity of gauged $\mathcal{N}=8$ supergravity.

It would be interesting to investigate which kind of instability is at work. There are at least two candidates: a Gregory-Lafflamme type of instability, due to the presence of flat directions; a superradiance effect that could appear in rotating black holes, as in our case, from the point of view of 11D supergravity, the system corresponds to rotating M2-branes along the $U(1)^4$ isometries of the $S^7$.
\\
We also considered 4-charge magnetic black hole configurations with different horizon topologies constructed in \cite{Duff:1999gh,Cvetic:1999xp}, and presented them in a slightly different form that allows us to prove that they do have a regular black hole BPS limit. This result shows that, when the embedding in the maximal theory exists, the non-extremal configurations which generalize \cite{Cacciatori:2009iz}, correspond to the black holes discussed in \cite{Duff:1999gh,Cvetic:1999xp} and, in the spherical case, they coincide with the non-extremal black holes presented in \cite{Toldo:2012ec}.
\\
Within the context of extremal magnetically charged black holes with flat horizon geometry, we identified three families of extremal black holes classified by their boundary conditions. One of them corresponds to the family of extremal BPS black holes, whose fields can be written as a solution of a first-order system of equations controlled by a superpotential \cite{DallAgata:2010ejj}, and the quartic invariant in the charges turns out to be negative. The thermodynamic stability analysis of the Hessian matrix imposing the topological twist condition ab initio indicates these configurations are metastable. The remaining two families of black holes are extremal non-BPS and they differ in the sign of the quartic invariant; this suggests that the sign of the quartic invariant is not a sufficient condition to identify a black hole configuration as supersymmetric or not. One of these families is obtained from the extremal BPS family by flipping the sign of the magnetic charge of the graviphoton, and thus satisfies a topological twist condition with one sign flipped. This class is also described by a first-order system controlled by a fake-superpotential \cite{Gnecchi:2012kb}. We assert thermodynamic quasi-stability for the black holes belonging to this family. The last family corresponds to extremal non-BPS configurations that exist for certain choices of boundary conditions in a complementary region in the space of the magnetic charges where the two latter families exist. These black holes are not described by a first-order system and its Hessian has generically negative eigenvalues indicating thermodynamic instability. In general, all these instabilities open the question of what is the phase diagram relevant for supergravity and its dual field theory, something that we expect to explore in the future along the lines of \cite{Anabalon:2024lgp}.


\section*{Acknowledgements}
The authors would like to thank to Roberto Emparan, Adolfo Guarino and David Mateos for useful discussion and comments. The work of AA is supported in part by the FONDECYT grants 1210635, 1221504, 1230853 and 1242043. MO is partially funded by Beca ANID de Doctorado grant 21222264.

\newpage

\appendix

\section{The STU model from maximal  supergravity}
\label{Appendix STU model}
{Let us recall in this appendix the basic facts about the STU model and its embeddings in the ${\rm SO}(8)$-gauged maximal supergravity \cite{Duff:1999gh}. This STU model describes $\mathcal{N}=2$ supergravity coupled to 3 vector multiplet, with a suitable Fayet-Iliopoulos (FI) term.
The four vector fields gauge a Cartan subalgebra of the $\mathfrak{so}(8)$ gauge algebra. The corresponding gauge group ${\rm SO}(2)^4={\rm SO}(2)_0\times{\rm SO}(2)_1\times{\rm SO}(2)_2\times{\rm SO}(2)_3 \subset {\rm SO}(8)$ can be chosen so that each factor act on one of the four couples $(A_I)$, $I=0,\dots, 3$:
\[
A_{0}=(1,2);\hspace{0.5cm}A_{1}=(3,4);\hspace{0.5cm}A_{2}=(5,6);\hspace{0.5cm}A_{3}=(7,8).
\]
in which the R-symmetry index of the eight gravitini $\psi_{{\tt i}\mu}$, ${\tt i}=1,\dots, 8$, of the maximal theory, can be split:
$$\Psi_{{\tt i}\mu}=\{\Psi_{A_I\,\mu}\}_{I=0,\dots,3}\,,$$
so that the couple of gravitini $\psi_{A_I\,\mu}$ transform as a doublet under ${\rm SO}(2)_I$.
The bosonic sector of the STU model is defined by the fields of the $\mathcal{N}=8$ theory which are singlets under ${\rm SO}(2)^4$. 
The 28 gauge fields $A_{\mu}^{{\tt IJ}}$, ${\tt I},{\tt J}=1,\dots 8$, of the parent model, transforming in the ${\bf 28}$ of ${\rm SL}(8,\mathbb{R})$, together with their magnetic duals $A_{{\tt IJ}\,\mu}$, yield, upon truncation, the following singlet vector fields 
\begin{align}
(A^M_\mu)=\boldsymbol{\Omega}\cdot (A^{12}_\mu,\,A^{34}_\mu,\,A^{56}_\mu,\,A^{78}_\mu,\,A_{12\,\mu},\,A_{34\,\mu},\,A_{56\,\mu},\,A_{78\,\mu})\,.
\end{align}
The above fields transform as a symplectic vector in the $({\bf 2},{\bf 2},{\bf 2})$ of the classical global symmetry group ${\rm SU}(1,1)^3$ of the $\mathcal{N}=2$ theory and $\boldsymbol{\Omega}$ is a constant  $8\times 8$ symplectic matrix encoding the freedom in the choice of the basis of the symplectic representation (\emph{symplectic frame}). We shall discuss the relevant symplectic frames below.\par
The 70 scalar fields $\phi^{{\tt ijkl}}$ of the maximal theory, transforming in the ${\bf 70}$
of ${\rm SU}(8)$,\footnote{Recall that the $\phi^{{\tt ijkl}}$ satisfy the reality condition $\phi^{\tt ijkl}=\frac{1}{24}\,\epsilon^{\tt ijkli'j'k'l'}\,(\phi^{\tt i'j'k'l'})^*$.} upon reduction to ${\rm SO}(2)^4$-singlets, reduce to three complex scalars
\begin{align*}
(\phi_{(12)(34)},\phi_{(12)(56)},\phi_{(12)(78)})=z^{i}=\chi_{i}+ie^{-\phi_{i}}\,,\,i=1,2,3
\end{align*}
spanning the scalar manifold of the STU model:
$$\mathscr{M}_{{\rm scal.}}=\left(\frac{{\rm SU}(1,1)}{{\rm U}(1)}\right)^3\,.$$
This is a \emph{special K\"ahler} manifold\footnote{See \cite{Trigiante:2016mnt}, and references therein, for a review of special geometry and of the gauging of $\mathcal{N}=2$ models in the embedding tensor formalism.} whose isotropy group ${\rm U}(1)^3$ being the subgroup of ${\rm SU}(8)$ commuting with the ${\rm SO}(2)^4$ residual gauge group. The isometry group ${\rm SU}(1,1)^3$ defines the on-shell global symmetry group of the classical theory, acting on the eight $A^M_\mu$ in the $({\bf 2},{\bf 2},{\bf 2})$ symplectic representation, as mentioned above.\par
The fermionic sector, on the other hand, depends on the factor ${\rm SO}(2)_I$, within ${\rm SO}(2)^4$, which is chosen to lie within the R-symmetry group of the $\mathcal{N}=2$ model.
This sector of the corresponding STU model, which will be referred to as ${\rm STU}_I$, is then defined as the singlet sector, within the maximal theory, with respect to the remaining group ${\rm SO}(2)^3={\rm SO}(2)_J\times {\rm SO}(2)_K\times {\rm SO}(2)_L$, $I\neq J\neq K\neq L\neq I$ and $\psi_{A_I\,\mu}$ are the spin-$3/2$ fields of the truncation. 
The consistency of this truncation is guaranteed by the fact that the STU$_I$ model, so defined, is the singlet sector with respect to the subgroup ${\rm SO}(2)^3={\rm SO}(2)_J\times {\rm SO}(2)_K\times {\rm SO}(2)_L$ of ${\rm SO}(8)$.
Choosing, for instance, the STU$_0$ model, whose gravitini are $\psi_{A_0\,\mu}=(\psi_{1\,\mu},\,\psi_{2\,\mu})$, the spin-$1/2$ fields which are singlets with respect to ${\rm SO}(2)_1\times {\rm SO}(2)_2\times{\rm SO}(2)_3$, are:
\begin{align*}
(\lambda^{A_{0}34},\lambda^{A_{0}56},\lambda^{A_{0}78})\equiv\lambda^{i A_{0}}\,,\quad i=1,2,3,\,A_{0}=1,2.
\end{align*}
and enter the three vector multiplets together with the complex scalar fields and three of the vector fields.\par
Upon reducing the gauge group of the maximal theory to ${\rm SO}(2)^4$ and the electric/magnetic duality index to the eight ${\rm SO}(2)^4$-singlets, the gauge connection reduces to:\footnote{We absorb the gauge grouping constant $g$ in the embedding tensor.}
\begin{equation}
A^M_\mu\,X_M=A^M_\mu\,\Theta_M{}^I\,t_I\,,
\end{equation}
where $t_I$ are the four generators of ${\rm SO}(2)^4$. In the STU$_I$ model, only $t_I$
has a non-trivial action on the fermionic fields and thus the only minimal couplings involve the spin-$1/2$ and $3/2$ fields and the single vector combination $A^M_\mu\,\Theta_M{}^I$. The 8-component symplectic vector $\Theta_M{}^I$ defines the FI term of the model and the corresponding scalar potential. For the sake of concreteness, we shall work in the STU$_0$ model, which we shall simply refer to as the STU model, denoted by $\theta_M$ the corresponding FI term $\Theta_M{}^0$ and by $A,B=1,2$ the R-symmetry indices $A_0,B_0$.\\
The geometry of the spacial K\"ahler manifold $\mathscr{M}_{\rm scal.}$ is described in terms of a holomorphic symplectic section $\Omega^M(z_i)$ which, in the symplectic frame that we adopt and modulo multiplication by a non-vanishing holomorphic function, reads:
\begin{align}
\Omega^{M} & =(z_{2}z_{3},z_{1}z_{3},z_{1}z_{2},-1,z_{1},z_{2},z_{3},-z_{1}z_{2}z_{3})\,,\label{Omegasf}
\end{align}
The K\"ahler potential, for instance, reads:
\begin{equation}
\mathcal{K}(z,\bar{z})=-\log(i\,\overline{\Omega}^T\mathbb{C}\,\Omega)=8\,{\rm Im}(z_1){\rm Im}(z_2){\rm Im}(z_3)=8\,e^{-\phi_1-\phi_2-\phi_3}\,.\label{Kpot}
\end{equation}
where
 $\bC$ is the ${\rm Sp}(8,\mathbb{R})$-invariant matrix:
 \begin{equation}
\bC=\left(\begin{array}{cc}
\mathbf{0} & \mathbf{1}\\
-\mathbf{1} & \mathbf{0}
\end{array}\right)
\end{equation}
The metric reads:
\begin{equation}
g_{i\bar{\jmath}}=\partial_i\partial_{\bar{\jmath}}\mathcal{K}=-\frac{1}{(z_i-\bar{z}_{\bar{\imath}})^2}\delta_{i\bar{\jmath}}=\sum_{\hat{\imath}=1}^3e_i^{\hat{\imath}}(e_j^{\hat{\imath}})^*\,,   \label{gijb} 
\end{equation}
where $e_i^{\hat{\imath}}$ is the complex vielbein matrix.
We also define the covariantly holomorphic symplectic section $V^M(z,\bar{z})\equiv e^{\mathcal{K}/2}\,\Omega(z)^M$, in terms of which the $\mathcal{N}=2$ central charge $\mathcal{Z}$ on a given solution and the gauge superpotential $\mathcal{W}$ induced by the FI term, read:
\begin{equation}
    \mathcal{Z}\equiv \Gamma^T\cdot\mathbb{C}\cdot V\,\,,\,\,\,\mathcal{W}\equiv V^T\cdot\theta\,,
\end{equation}
$\Gamma=(\Gamma^M)=(P^\Lambda,\,-Q_\Lambda)$ being the symplectic vector of quantized charges.\par
The relationship between the different STU$_I$ models within the maximal one can be inferred from inspection of the gravitino shift-matrix ${\rm A}_{1{\tt ij}}$ as a function of the singlet scalars, whose only non-vanishing entries are:
\begin{align}
{\rm A}_{1\,A_0,A_0}\,(L^{-1})&=\frac{1}{\sqrt{2}}\,\overline{\mathcal{W}}=\frac{1}{\sqrt{2}}\,\overline{V}^T\cdot\theta\,,\nonumber\\
{\rm A}_{1\,A_i,A_i}\,(L^{-1})&=\frac{1}{\sqrt{2}}\,\overline{\mathcal{W}}_{(i)}=\frac{1}{\sqrt{2}}\,\mathcal{D}_{\hat{\imath}}V^M\,\Theta_M{}^i\,,\,\,\,i=1,2,3\,,
\end{align}
where $\mathcal{D}_iV^M$ are the K\"ahler-covariant derivatives of $V^M$: 
$$\mathcal{D}_{\hat{\imath}}V^M\equiv e_{\hat{\imath}}^i\,\mathcal{D}_{i}V^M=e_{\hat{\imath}}^i\,\left(\partial_i+\frac{1}{2}\,\partial_i\mathcal{K}\right)V^M\,\,,\,\,e_{\hat{\imath}}^i=-(z_i-\bar{z}_{\bar{\imath}})\delta_{\hat{\imath}}^i\,.$$
If $\mathcal{W}_{(0)}\equiv \mathcal{W}$ is the gauge superpotential in the STU=STU$_0$ model, $\mathcal{W}_{(i)}$ is the corresponding function in the STU$_i$ model.
In the chosen symplectic frame, we have:
\begin{align}
    \theta_M&\equiv \Theta_M{}^0=\frac{1}{\sqrt{2}L}\,(1,1,1,1,0,0,0,0)\,,\label{embedding tensor}\\
    \Theta_M{}^1&=\mathcal{O}_{1\,M}{}^N\,  \theta_N\,,\,\, \Theta_M{}^2=\mathcal{O}_{2\,M}{}^N\,\theta_N\,,\,\,\Theta_M{}^3=\mathcal{O}_{3\,M}{}^N\,\theta_N\,,\nonumber
\end{align}
where the symplectic matrices $\mathcal{O}_i$, $i=1,2,3$ read:\footnote{The matrices $\mathcal{O}_{i} $ ($i=1,2,3$),together with the identity matrix $\mathbb{1}_{8\times8}$ define a Klein group $\mathbb{Z}_2\times \mathbb{Z}_2$. They satisfy the relation
$$\mathcal{O}_{i}\cdot\mathcal{O}_{j} =\delta_{ij}\,\mathbb{1}+|\epsilon_{ijk}| \mathcal{O}_{k}\,.$$
}
\begin{align}
    \mathcal{O}_1&={\rm diag}(1,-1,-1,1,1,-1,-1,1)\,,\nonumber\\
    \mathcal{O}_2&={\rm diag}(-1,1,-1,1,-1,1,-1,1)\,,\nonumber\\
    \mathcal{O}_3&={\rm diag}(-1,-1,1,1,-1,-1,1,1)\,.\nonumber
\end{align}
Moreover, one can verify that:
\begin{align}
    \mathcal{O}_i\cdot V(z,\bar{z})&=\left.\overline{\mathcal{D}}_{\hat{\imath}}\overline{V}\right\vert_{(z_i\rightarrow z_i\,,\,\,z_{j\neq i}\rightarrow -\bar{z}_{\bar{\jmath}})}\,,\nonumber\\
     \mathcal{O}_i\cdot {\mathcal{D}}_{\hat{\jmath}}{V}&=|\epsilon_{ijk}|\left.{\mathcal{D}}_{\hat{k}}{V}\right\vert_{(z_i\rightarrow z_i\,,\,\,z_{j\neq i}\rightarrow -\bar{z}_{\bar{\jmath}})}\,.
\end{align}
Using the first of the above properties, one can verify that:
\begin{equation}
\mathcal{W}_{(i)}=\left.\overline{\mathcal{W}}\right\vert_{(z_i\rightarrow z_i\,,\,\,z_{j\neq i}\rightarrow -\bar{z}_{\bar{\jmath}})}\,.
\end{equation}
We then conclude that solutions to the STU =STU$_0$ model are mapped to solutions of the STU$_i$ one through the following transformation:
\begin{align}
\mathrm{STU}_0\rightarrow\mathrm{STU}_i\,\Leftrightarrow\,\,\begin{cases}\Gamma\rightarrow \mathcal{O}_i\cdot\Gamma\,, & \theta\rightarrow \Theta^i=\mathcal{O}_i\cdot\theta
\cr z_i\rightarrow z_i\,, & \,z_{j\neq i}\rightarrow -\bar{z}_{\bar{\jmath}}\end{cases} 
\end{align}
Under this transformation, a BPS solution of the STU model is mapped into a BPS solution of the STU$_i$ one, which is non-BPS in the original truncation but BPS in the maximal theory. The action of $\mathcal{O}_i$ for going from STU$_0$ to STU$_i$, at the level of R-symmetry indices, as an ${\rm SU}(8)$ compensating transformation, has the effect to exchanging the couples $(A_0)=(1,2)$ with $(A_i)$ and $(A_j)$ with  $(A_k)$, $i\neq j\neq k\neq i$. For instance, $\mathcal{O}_1$ implies $(1,2)\leftrightarrow (3,4)$ and $(5,6)\leftrightarrow (7,8)$.
} 
\paragraph{Action and supersymmetry transformations}

The $\mathcal{N}=2$ gauged STU model has the following bosonic action
principle (we set $8\pi G_N=1$):

\begin{align}
\mathcal{S} & =\int\dd^{4}x\sqrt{-g}\left(\frac{R}{2}-g_{i\overline{\jj}}\partial_{\mu}z^{i}\partial^{\mu}\overline{z}^{\overline{\jj}}-V(z,\overline{z})+\frac{1}{4}\mathcal{I}_{\Lambda\Sigma}F_{\mu\nu}^{\Lambda}F^{\Sigma\mu\nu}+\frac{1}{8e}\mathcal{R}_{\Lambda\Sigma}F_{\mu\nu}^{\Lambda}F_{\rho\sigma}^{\Sigma}\varepsilon^{\mu\nu\rho\sigma}\right)\,.\nonumber \\
\label{action STU-1-1}
\end{align}
where $i,j=1,2,3$, $\Lambda,\Sigma=1,\dots,4$ and we are considering
the symplectic frame described by a holomorphic section of the form (\ref{Omegasf}).
This frame is associated with a prepotential of the form 
\begin{align}
F(X) & =2i\sqrt{X^{0}X^{1}X^{2}X^{3}}\,.
\end{align}
Indeed $\Omega^M(z)$, modulo multiplication by a non-vanishing holomorphic function, can be written in the form $\Omega^{M}=(X^{\Lambda},\partial F/\partial X^\Lambda)$ provided we identify:
\begin{align}
z_{1}=-i\sqrt{\frac{X^{1}X^{2}}{X^{0}X^{3}}}\,,\quad z_{2}=-i\sqrt{\frac{X^{0}X^{2}}{X^{1}X^{3}}}\,,\quad z_{3}=-i\sqrt{\frac{X^{0}X^{1}}{X^{2}X^{3}}}\,.
\end{align}
The K\"ahler potential and the metric are given in eqs \eqref{Kpot} and \eqref{gijb}, respectively.\par
The supersymmetry variations of the fermions in a bosonic background
are given by\footnote{We consider $(\sigma^{2})^{A}{}_{C}=-i\delta_{1}^{A}\delta_{C}^{2}+i\delta_{2}^{A}\delta_{C}^{1}$,
for which the following identities hold $i(\sigma^{2})^{A}{}_{C}\varepsilon^{BC}=\delta^{AB}$,
$i(\sigma^{2})^{A}{}_{B}\epsilon^{B}=\varepsilon^{AC}\delta_{CB}\epsilon^{B}$,
$i(\sigma^{2})_{C}{}^{B}\varepsilon^{CA}=\delta^{AB}$. Requiring
that $(\sigma^{2})^{1}{}_{2}=(\sigma^{2})_{1}{}^{2}$, meaning that
they are the same matrices, implies that $(\sigma^{2})_{B}{}^{A}=-i\delta_{B}^{1}\delta_{2}^{A}+i\delta_{B}^{2}\delta_{1}^{A}$
leading to the identity $i(\sigma^{2})_{C}{}^{B}\varepsilon^{CA}=\delta^{AB}$.}
\begin{align}
\delta\Psi_{\mu}^{A} & =D_{\mu}\epsilon^{A}+\frac{1}{4}L^{\Lambda}\mathcal{I}_{\Lambda\Sigma}F_{\rho\sigma}^{\Sigma+}\gamma^{\rho\sigma}\gamma_{\mu}\varepsilon^{AB}\epsilon_{B}+\frac{1}{2}\underbrace{i(\sigma^{2})^{A}{}_{C}\varepsilon^{BC}}_{\delta^{AB}}\mathcal{W}\gamma_{\mu}\epsilon_{B}\,,\label{general susy var 1}\\
\delta\lambda^{iA} & =-\partial_{\mu}z^{i}\gamma^{\mu}\epsilon^{A}+\frac{1}{2}g^{i\overline{\jj}}\overline{f}_{\overline{\jj}}^{\Lambda}\mathcal{I}_{\Lambda\Sigma}F_{\mu\nu}^{\Sigma-}\gamma^{\mu\nu}\varepsilon^{AB}\epsilon_{B}+W^{iAB}\epsilon_{B}\,.\label{general susy var 2}
\end{align}
where 
\begin{align}
D_{\mu}\epsilon^{A} & =\partial_{\mu}\epsilon^{A}+\frac{1}{4}\omega_{\mu}{}^{ab}\gamma_{ab}\epsilon^{A}+\frac{1}{2}A_{\mu}^{M}\theta_{M}\underbrace{i(\sigma^{2})^{A}{}_{B}\epsilon^{B}}_{\varepsilon^{AB}\delta_{BC}\epsilon^{C}}+\frac{i}{2}\mathcal{Q}_{\mu}\epsilon^{A}\,,\\
\mathcal{Q}_{\mu} & =\frac{i}{2}(\partial_{\overline{\ii}}\mathcal{K}\partial_{\mu}\overline{z}^{\overline{\ii}}-\partial_{i}\mathcal{K}\partial_{\mu}z^{i})=\frac{1}{2}\sum_{i=1}^{3}e^{\phi_{i}}\partial_{\mu}\chi_{i}\,,\\
V^{M} & =e^{\mathcal{K}/2}\Omega^{M}=(L^{\Lambda},M_{\Lambda})\,,\quad\mathcal{K}=-\log[-i\Omega(z)^{T}\bC\overline{\Omega}(\overline{z})]\,,\\
\mathcal{W} & =V^{M}\theta_{M}\,,\\
V(z,\overline{z}) & =g^{i\overline{\jj}}\mathscr{D}_{i}\mathcal{W}\overline{\mathscr{D}}_{\overline{\jj}}\mathcal{W}-3\mathcal{W}\overline{\mathcal{W}}\,,\\
W^{iAB} & =\underbrace{i(\sigma^{2})_{C}{}^{B}\varepsilon^{CA}}_{\delta^{AB}}g^{i\overline{\jj}}\overline{\mathscr{D}}_{\overline{\jj}}(\overline{V}^{M}\theta_{M})\,.
\end{align}
and the embedding tensor $\theta_M$ was given in \eqref{embedding tensor}. 
\paragraph{Relation to other symplectic frames.}
Let us now give the explicit relation between the symplectic frame used in the present work and other frames commonly used in the literature.\par
\emph{Frame 1.} The first is the symplectic frame, which we shall refer to as \emph{cubic frame}, in which the prepotential function $\tilde{F}(\tilde{X})$ has the following cubic form:
\begin{equation}
    \tilde{F}(\tilde{X})=\frac{d_{ijk}}{3!}\,\frac{\tilde{X}^i \tilde{X}^j \tilde{X}^k}{\tilde{X}^0}=-\frac{\tilde{X}^1 \tilde{X}^2 \tilde{X}^3}{\tilde{X}^0}\,.
\end{equation}
The corresponding holomorphic section $\tilde{\Omega}(z)=(\tilde{X}^\Lambda,\,\partial \tilde{F}/\partial \tilde{X}^\Lambda)$ reads, modulo multiplication by a non-vanishing holomorphic function:
\begin{equation}
    \tilde{\Omega}(z)=(1,z_1,z_2,z_3,z_1 z_2 z_3,-z_2 z_3,-z_1 z_3,-z_1 z_2)\,,
\end{equation}
where 
$$z_i=\frac{\tilde{X}^i}{\tilde{X}^0}=\chi_i+i\,e^{-\phi_i}\,\,,\,\,\,i=1,2,3\,.$$
The two frames are related by the following symplectic matrix ${\bf E}=(E^M{}_N)$:
\begin{align}
    \Omega(z)&={\bf E}\cdot \tilde{\Omega}(z)\,\,,\,\,\,\,
    {\bf E}=\left(
\begin{array}{cccccccc}
 0 & 0 & 0 & 0 & 0 & -1 & 0 & 0 \\
 0 & 0 & 0 & 0 & 0 & 0 & -1 & 0 \\
 0 & 0 & 0 & 0 & 0 & 0 & 0 & -1 \\
 -1 & 0 & 0 & 0 & 0 & 0 & 0 & 0 \\
 0 & 1 & 0 & 0 & 0 & 0 & 0 & 0 \\
 0 & 0 & 1 & 0 & 0 & 0 & 0 & 0 \\
 0 & 0 & 0 & 1 & 0 & 0 & 0 & 0 \\
 0 & 0 & 0 & 0 & -1 & 0 & 0 & 0 \\
\end{array}
\right)\,.
\end{align}
The charges $\Gamma^M=(P^\Lambda,\,Q_\Lambda)$ in our frame are then related to those $\tilde{\Gamma}^M=(\mathfrak{p}^\Lambda,\,\mathfrak{q}_\Lambda)$ as follows:
\begin{equation}
    \Gamma^M=(-\mathfrak{q}_1, -\mathfrak{q}_2, -\mathfrak{q}_3, -\mathfrak{p}^0, \mathfrak{p}^1, \mathfrak{p}^2, \mathfrak{p}^3, -\mathfrak{q}_0)\,.\label{cubictostandard}
    \end{equation}
The quartic invariant, see Appendix \ref{appendix quartic invariant}, in the cubic frame, reads:
\begin{equation}
    I_4(\tilde{\Gamma})=-(\mathfrak{p}^\Lambda \mathfrak{q}_\Lambda)^2-4\,\mathfrak{q}_0\,\mathfrak{p}^1\,\mathfrak{p}^2\,\mathfrak{p}^3+4\,\mathfrak{p}^0\,\mathfrak{q}_1\,\mathfrak{q}_2\,\mathfrak{q}_3+4\,\left(\sum_{i<j}\mathfrak{p}^i \mathfrak{q}_i\mathfrak{p}^j \mathfrak{q}_j\right)\,.
\end{equation}
In light of eq. \eqref{cubictostandard}, we can write the quartic invariant in our frame, in the magnetic case $Q_\Lambda=0$, in terms of the charges in the cubic one as follows
$$I_4(\Gamma)=4 P^1 P^2 P^3 P^4=4 \mathfrak{p}^0\,\mathfrak{q}_1\,\mathfrak{q}_2\,\mathfrak{q}_3\,.$$
\par
\emph{Frame 2.} The second symplectic frame is the one which naturally arises from direct truncation of the ${\rm SO}(8)$ gauged maximal theory. It is a special coordinate frame with a prepotential function $\hat{F}(\hat{X})$ of the form:
\begin{equation}
    \hat{F}(\hat{X})=-2\sqrt{\hat{X}^0\hat{X}^1\hat{X}^2\hat{X}^3 }\,.
\end{equation}
The holomorphic section $\hat{\Omega}=(\hat{X}^\Lambda,\,\partial \hat{F}/\partial \hat{X}^\Lambda)$, modulo multiplication by a non-vanishing holomorphic function, can be written in the form:
\begin{equation}
    \hat{\Omega}(z)=(z_1 z_2 z_3,z_1,z_2,z_3,-1,-z_2 z_3,-z_1 z_3,-z_1 z_2)\,,
\end{equation}
where:
\begin{equation}
    z_i=\sqrt{\frac{\hat{X}^0 \hat{X}^i}{\hat{X}^j\hat{X}^k}}\,\,,\,\,\,i\neq j\neq k\neq i\,.
\end{equation}
The symplectic transformation relating this frame with the cubic one is straightforward:
$$\hat{X}^0=\frac{\partial \tilde{F}}{\partial \tilde{X}^0}\,,\,\,\hat{X}^i=\tilde{X}^i\,,\,\,\frac{\partial \hat{F}}{\partial \hat{X}^0}=-\tilde{X}^0\,,\,\,\frac{\partial \hat{F}}{\partial \hat{X}^i}=\frac{\partial \tilde{F}}{\partial \tilde{X}^i}\,.$$

\section{Spinor conventions}
We use the Majorana basis for the Clifford algebra
\begin{align}
\gamma^{0}=-i\left(\begin{array}{cc}
0 & \sigma_{2}\\
\sigma_{2} & 0
\end{array}\right)\,,\quad\gamma^{1}=-\left(\begin{array}{cc}
\sigma_{3} & 0\\
0 & \sigma_{3}
\end{array}\right)\,,\quad\gamma^{2}=i\left(\begin{array}{cc}
0 & -\sigma_{2}\\
\sigma_{2} & 0
\end{array}\right)\,,\quad\gamma^{3}=\left(\begin{array}{cc}
\sigma_{1} & 0\\
0 & \sigma_{1}
\end{array}\right)\,.
\end{align}
The charge conjugation matrix and $\gamma^{5}$ matrix are given by
\begin{align}
C & =\gamma_{0}\,, & \gamma^{5} & =i\gamma^{0}\gamma^{1}\gamma^{2}\gamma^{3}\,.
\end{align}
We use $\mathcal{N}=2$ chiral supersymmetry parameters $\epsilon^{A},\epsilon_{A}$
with $A=1,2$, satisfying
\begin{align}
\gamma^{5}\epsilon^{A} & =-\epsilon^{A}\,, & \gamma^{5}\epsilon_{A} & =\epsilon_{A}\,,
\end{align}
that are defined as the chiral components of doublet of Majorana spinors $\psi^{A}=\epsilon^{A}+\epsilon_{A}$.
The relation between the chiral spinors is $\epsilon_{A}=(\epsilon^{A})^{*}$.
It is useful to define the complex spinors 
\begin{align}
\zeta = \psi^1+\ri \psi^2
\end{align}
that allows one to write down a set of differential equations. The general rule to go from an equation with real coefficients for the Majorana spinor $\psi^A$ to an equation for the complex spinor $\zeta$ is by replacing $\delta_{AB}\to 1$ and $\varepsilon_{A B}\to -\ri$, and vice-versa. 

\section{Dirac bracket between supercharges}\label{Appendix Dirac bracket}
The author of \cite{Hristov:2011qr} showed that the Dirac bracket between
the supercharges is expressed as 
\begin{align}
\{\mathcal{Q},\mathcal{Q}\} & =\int_{\partial\hM}\dd\Sigma_{\mu\nu}\epsilon^{\mu\nu\rho\sigma}\left(\overline{\epsilon}^{A}\gamma_{\rho}\mathfrak{D}_{A\sigma}(\epsilon)-\overline{\epsilon}_{A}\gamma_{\rho}\mathfrak{D}^{A}{}_{\sigma}(\epsilon)\right)\,,\label{bracket between supercharges}
\end{align}
where $\mathfrak{D}^{A}{}_{\sigma}(\epsilon)$ is generically defined
by the variation of the gravitino $\Psi_{\sigma}^{A}$. Since we
are interested in the purely dilatonic sector, $\mathfrak{D}^{A}{}_{\sigma}(\epsilon)$
is given by \eqref{KS equation chiral} and $\mathfrak{D}_{A\sigma}(\epsilon)$
is its conjugated, explicitly 
\begin{align}
\mathfrak{D}_{A}(\epsilon) & =\dd\epsilon_{A}+\frac{1}{4}\omega_{ab}\gamma^{ab}\epsilon_{A}+\frac{1}{2}A^{M}\theta_{M}\varepsilon_{AB}\delta^{BC}\epsilon_{C}+\frac{1}{4}\overline{L}^{T}\mathcal{I}F_{ab}\gamma^{ab}\gamma\varepsilon_{AB}\epsilon^{B}+\frac{1}{2}\overline{\mathcal{W}}\gamma\delta_{AB}\epsilon^{B}\,,
\end{align}
we used the fact that the K\"ahler connection vanishes. The 2-form volume
is defined as
\begin{align}
\dd\Sigma_{\mu\nu} & =\frac{1}{2}\epsilon_{\mu\nu\rho\sigma}\dd x^{\rho}\wedge\dd x^{\sigma}\,.
\end{align}
First, we simplify the contraction appearing in the integral, namely
\begin{align}
\dd\Sigma_{\mu\nu}\epsilon^{\mu\nu\rho\sigma} & =\frac{1}{2}\epsilon_{\mu\nu\lambda\delta}\epsilon^{\mu\nu\rho\sigma}\dd x^{\lambda}\wedge\dd x^{\delta}\,.\label{sigma epsilon}
\end{align}
Considering the definitions of the symbolic with curved indices
\begin{align}
\epsilon_{\mu\nu\rho\sigma} & =e^{-1}e^{a}{}_{\mu}e^{b}{}_{\nu}e^{c}{}_{\rho}e^{d}{}_{\sigma}\epsilon_{abcd}\,,\hspace{1cm}\epsilon^{\mu\nu\rho\sigma}=ee_{a}{}^{\mu}e_{b}{}^{\nu}e_{c}{}^{\nu}e_{d}{}^{\rho}\epsilon^{abcd}\,,
\end{align}
where the symbol with flat indices is $\epsilon_{0123}=1=-\epsilon^{0123}$.
Hence,
\begin{align}
\epsilon_{\mu\nu\lambda\delta}\epsilon^{\mu\nu\rho\sigma} & =e^{c}{}_{\lambda}e^{d}{}_{\delta}e_{g}{}^{\rho}e_{h}{}^{\sigma}\epsilon_{abcd}\epsilon^{abgh}\,.
\end{align}
using the identity $\epsilon_{abcd}\epsilon^{abgh}=-4\delta_{[c}^{g}\delta_{d]}^{h}$,
it follows that $\epsilon_{\mu\nu\lambda\delta}\epsilon^{\mu\nu\rho\sigma}=-2(\delta_{\lambda}^{\rho}\delta_{\delta}^{\sigma}-\delta_{\lambda}^{\sigma}\delta_{\delta}^{\rho})$,
which leads to the following simplified form of (\ref{sigma epsilon})
\begin{align}
\dd\Sigma_{\mu\nu}\epsilon^{\mu\nu\rho\sigma} & =-(\delta_{\lambda}^{\rho}\delta_{\delta}^{\sigma}-\delta_{\lambda}^{\sigma}\delta_{\delta}^{\rho})\dd x^{\lambda}\wedge\dd x^{\delta}=-2\dd x^{\rho}\wedge\dd x^{\sigma}\,.
\end{align}
Replacing this result into the bracket between supercharges (\ref{bracket between supercharges})
we obtain \eqref{backet between call Q}. The expression \eqref{backet between call Q}
can be obtained by defining a Majorana spinor 
\begin{align}
\psi^{A} & =\epsilon_{A}+\epsilon^{A}
\end{align}
which implies the following relations
\begin{align}
\epsilon_{A} & =\mathtt{P}_{+}\psi^{A}\,,\hspace{1.5cm}\epsilon^{A}=\mathtt{P}_{-}\psi^{A}\,,\hspace{1.5cm}\mathtt{P}_{\pm}=\frac{1}{2}(1\pm\gamma_{5})\,.
\end{align}
We recall that the conjugated of the chiral spinors is defined as
$\overline{\epsilon}_{A}=\ri(\epsilon^{A})^{\dagger}\gamma^{0}$ and
$\overline{\epsilon}^{A}=\ri(\epsilon_{A})^{\dagger}\gamma^{0}$,
in order to preserve the chirality. In our basis $\gamma_{5}^{\dagger}=\gamma_{5}$, so one can check that
\begin{align}
\overline{\epsilon}_{A} & =\overline{\psi}^{A}\mathtt{P}_{+}\,,\hspace{1.5cm}\overline{\epsilon}^{A}=\overline{\psi}^{A}\mathtt{P}_{-}\,.
\end{align}
Now, we use the fact that the electric components of the U(1) section
and the superpotential are real functions, i.e. $L^{\Lambda},\mathcal{W}\in\bR$.
Then, the supercovariant derivatives $\mathfrak{D}^{A}(\epsilon),\mathfrak{D}_{A}(\epsilon)$
can be written in terms of the Majorana spinor as follows 
\begin{align}
\mathfrak{D}^{A}(\epsilon) & =\mathtt{P}_{-}\dd\psi^{A}+\frac{1}{4}\omega_{ab}\gamma^{ab}\mathtt{P}_{-}\psi^{A}+\frac{1}{2}A^{M}\theta_{M}\varepsilon^{AB}\delta_{BC}\mathtt{P}_{-}\psi^{C}+\\
 & \hspace{1cm}+\frac{1}{4}L^{T}\mathcal{I}F_{ab}\gamma^{ab}\gamma\varepsilon^{AB}\mathtt{P}_{+}\psi^{B}+\frac{1}{2}\mathcal{W}\gamma\delta^{AB}\mathtt{P}_{+}\psi^{B}\,,\nonumber \\
\mathfrak{D}_{A}(\epsilon) & =\delta_{AB}\mathtt{P}_{+}\dd\psi^{B}+\frac{1}{4}\omega_{ab}\gamma^{ab}\delta_{AB}\mathtt{P}_{+}\psi^{B}+\frac{1}{2}A^{M}\theta_{M}\varepsilon_{AB}\mathtt{P}_{+}\psi^{B}+\\
 & \hspace{1cm}+\frac{1}{4}L^{T}\mathcal{I}F_{ab}\gamma^{ab}\gamma\varepsilon_{AB}\mathtt{P}_{-}\psi^{B}+\frac{1}{2}\mathcal{W}\gamma\delta_{AB}\mathtt{P}_{-}\psi^{B}\,,\nonumber 
\end{align}
The 2-form appearing in the integral of the Dirac bracket are
\begin{align*}
\overline{\epsilon}^{A}\gamma\wedge\mathfrak{D}_{A}(\epsilon) & =\overline{\psi}^{A}\gamma\wedge\mathtt{P}_{-}\left(\dd\psi^{A}+\frac{1}{4}\omega_{ab}\gamma^{ab}\psi^{A}+\frac{1}{2}A^{M}\theta_{M}\varepsilon^{AB}\delta_{BC}\psi^{C}+\right.\\
 & \hspace{1cm}\left.+\frac{1}{4}L^{T}\mathcal{I}F_{ab}\gamma^{ab}\gamma\varepsilon^{AB}\psi^{B}+\frac{1}{2}\mathcal{W}\gamma\delta^{AB}\psi^{B}\right)\,,\\
\overline{\epsilon}^{A}\gamma\wedge\mathfrak{D}_{A}(\epsilon) & =\overline{\psi}^{A}\gamma\mathtt{P}_{+}\wedge\left(\delta_{AB}\dd\psi^{B}+\frac{1}{4}\omega_{ab}\gamma^{ab}\delta_{AB}\psi^{B}+\frac{1}{2}A^{M}\theta_{M}\varepsilon_{AB}\psi^{B}+\right.\\
 & \hspace{1cm}\left.+\frac{1}{4}L^{T}\mathcal{I}F_{ab}\gamma^{ab}\gamma\varepsilon_{AB}\psi^{B}+\frac{1}{2}\mathcal{W}\gamma\delta_{AB}\psi^{B}\right)\,,
\end{align*}
Clearly, its subtraction cancels the identity factor in $\mathtt{P}_{\pm}$
leading to \eqref{backet between call Q}.

\section{Quartic invariant of $SL(2,\bR)^{3}$}

\label{appendix quartic invariant}

To construct the quartic invariants of $\mathrm{SL}(2,\bR){}^{3}\subset\mathrm{Sp}(8,\bR)$
we consider its generators given by
\begin{align}
\mathtt{X}_{i} & =\left.\frac{\partial\mathcal{M}}{\partial\chi_{i}}\right|_{\chi,\phi=0}\,,\hspace{1cm}\mathtt{Y}_{i}=\left.\frac{\partial\mathcal{M}}{\partial\phi_{i}}\right|_{\chi,\phi=0}\,,\hspace{1cm}\mathtt{Z}_{i}=[\mathtt{Y}_{i},\mathtt{X}_{i}]\,,\qquad(\text{no sum over }i)
\end{align}
with $i=1,2,3$. They span the three commuting $\mathfrak{sl}(2,\bR)$
algebras. It is convenient to consider the basis with nilpotent generators
$\mathtt{e}_{i}^{2}=\mathtt{f}_{i}^{2}=0$ and the generators $\mathtt{h}_{i}$
of the Cartan subalgebra, given by
\begin{align}
\mathtt{h}_{i}=\frac{1}{2}\mathtt{Y}_{i}\,,\hspace{1cm}\mathtt{e}_{i}=\frac{1}{4\sqrt{2}}(\mathtt{Z}_{i}+2\mathtt{X}_{i})\,,\hspace{1cm}\mathtt{f}_{i}=\mathtt{e}_{i}^{T}\,,\\{}
[\mathtt{h}_{i},\mathtt{e}_{i}]=\mathtt{e}_{i}\,,\hspace{1cm}[\mathtt{e}_{i},\mathtt{f}_{i}]=\mathtt{h}_{i}\,,\hspace{1cm}[\mathtt{h}_{i},\mathtt{f}_{i}]=-\mathtt{f}_{i}\,.\nonumber 
\end{align}
We collect all the generators as $t_{(i)\alpha}=\{\mathtt{e}_{i},\mathtt{h}_{i},\mathtt{f}_{i}\}$
with $\alpha=1,2,3$. By definition the positions of the symplectic
indices are $t_{(i)\alpha}=(t_{(i)\alpha})_{M}{}^{N}$and we lower
them by the symplectic matrix $\bC=\bC_{MN}$ defining $t_{(i)\alpha}\bC=(t_{(i)\alpha})_{MN}$.
We construct the Cartan-Killing form $\eta_{(i)\alpha\beta}=\tr(t_{(i)\alpha}t_{(i)\beta})$
and its inverse, denoted by $\eta_{(i)}^{\alpha\beta}$. These allow us
to define $(t_{(i)}^{\alpha})_{MN}=\eta_{(i)}^{\alpha\beta}(t_{(i)\beta})_{MN}$
and construct the following tensors of $\otimes^{4}\underline{8}$
\begin{align}
\mathcal{C}_{MNPQ} & =\bC_{MN}\bC_{PQ}\,,\\
(\mathcal{I}_{i})_{MNPQ} & =(t_{(i)}^{\alpha})_{MN}(t_{(i)\alpha})_{PQ}\,,\\
(\mathcal{L}_{ij})_{MNPQ} & =(t_{(i)}^{\alpha})_{M\bullet}(t_{(j)}^{\beta})^{\bullet}{}_{N}(t_{(i)\alpha})_{P\bullet}(t_{(j)\beta})^{\bullet}{}_{Q}\,,\\
(\mathcal{Z}_{ijk})_{MNPQ} & =(t_{(i)}^{\alpha})_{M\bullet}(t_{(j)}^{\beta})^{\bullet}{}_{\bullet}(t_{(k)}^{\gamma})^{\bullet}{}_{N}(t_{(i)\alpha})_{P\bullet}(t_{(j)\beta})^{\bullet}{}_{\bullet}(t_{(k)\gamma})^{\bullet}{}_{Q}\,.
\end{align}
which are invariant under $\mathrm{SL}(2,\bR)^{3}$. Among all the
above tensors only eight of them are independent, and one can pick
these to be $\{\mathcal{C},\mathcal{I}_{1},\mathcal{I}_{2},\mathcal{I}_{3},\mathcal{L}_{12},\mathcal{L}_{13},\mathcal{L}_{23},\mathcal{Z}_{123}\}$. Nevertheless, when these tensors act on two arbitrary symplectic vectors the eight invariants reduce to seven independent functions.

In our case we have two symplectic vectors $\Gamma^{M}$ and $\theta^{M}=\bC^{MN}\theta_{N}$,
hence we can construct the following independent invariants
\begin{align}
\mathcal{I}_{1}(\Gamma,\Gamma,\Gamma,\Gamma) & =-2\prod_{\Lambda}P_{\Lambda}\,,\\
\mathcal{I}_{1}(\theta,\theta,\theta,\theta) & =-\frac{1}{2L^{2}}\,,\\
\mathcal{I}_{i}(\Gamma,\theta,\Gamma,\theta) & =\frac{1}{16L^{2}}\left(\sum_{\Lambda}\Omega_{i\Lambda}P_{\Lambda}\right)^{2}\,.
\end{align}
The rest are zero or functionally dependent on the above ones.
Note that the BPS mass can be expressed in terms of the invariants as
\begin{align}
m_{\mathrm{BPS}}^{4}=64L^{3}\prod_{i}\mathcal{I}_{i}(\Gamma,\theta,\Gamma,\theta).
\end{align}
Other invariants that are dependent on the above are the following.
\begin{align}
\mathcal{C}(\Gamma,\theta,\Gamma,\theta) & =\frac{1}{2L^{2}}\left(\sum_{\Lambda}P_{\Lambda}\right)^{2}\,,\\
\mathcal{I}_{1}(\Gamma,\Gamma,\theta,\theta) & =-\frac{1}{2L^{2}}(P_{2}P_{3}+P_{1}P_{4})\,,\\
\mathcal{I}_{2}(\Gamma,\Gamma,\theta,\theta) & =-\frac{1}{2L^{2}}(P_{1}P_{3}+P_{2}P_{4})\,,\\
\mathcal{I}_{3}(\Gamma,\Gamma,\theta,\theta) & =-\frac{1}{2L^{2}}(P_{1}P_{2}+P_{3}P_{4})\,,\\
\mathcal{Z}_{123}(\Gamma,\Gamma,\theta,\theta) & =-\frac{1}{128L^{2}}\prod_{\Lambda\geq\Sigma}P_{\Lambda}P_{\Sigma}\,,
\end{align}
\hypersetup{linkcolor=blue}
\phantomsection 
\addtocontents{toc}{\protect\addvspace{4.5pt}}
\bibliographystyle{mybibstyle}
\bibliography{biblio.bib}

\end{document}